\def\version{April 4th, 1998}

\documentclass[reqno,11pt]{amsart}

\usepackage{epsf}

\def\be{\begin{equation}}
\def\ba{\begin{align}}
\def\bm{\begin{multline}}
\def\bfig{\begin{figure}[htb]}
\def\efig{\end{figure}}

\newcommand{\bibit}[1]{\vspace{1mm} \bibitem[#1]{#1}}
\newcommand{\paper}[1]{{\it #1}, }
\newcommand{\journal}[4]{#1 {\bf #2}, #3 (#4)}
\newcommand{\CMP}{Commun. Math. Phys.}
\newcommand{\HPA}{Helv. Phys. Acta}
\newcommand{\JSP}{J. Stat. Phys.}
\newcommand{\PR}{Phys. Rev.}
\newcommand{\PRA}{Phys. Rev. A}

\numberwithin{equation}{section}
\newtheorem{theorem}{Theorem}[section]

\newtheorem{lemma}{Lemma}[section]
\newtheorem{assumption}{Assumption}
\newtheorem*{property}{Property}

\newcommand{\cf}{c.f.\;}
\newcommand{\eg}{e.g.\;}
\newcommand{\ie}{i.e.\;}

\newcommand{\nn}{\nonumber}
\def\bbbone{{\mathchoice {\rm 1\mskip-4mu l} {\rm 1\mskip-4mu l}
{\rm 1\mskip-4.5mu l} {\rm 1\mskip-5mu l}}}

\DeclareMathSymbol{\leqslant}{\mathalpha}{AMSa}{"36}
\DeclareMathSymbol{\geqslant}{\mathalpha}{AMSa}{"3E}
\DeclareMathSymbol{\doteqdot}{\mathalpha}{AMSa}{"2B}
\DeclareMathSymbol{\circlearrowright}{\mathalpha}{AMSa}{"08}
\DeclareMathSymbol{\subsetneq}{\mathalpha}{AMSb}{"28}
\renewcommand{\leq}{\;\leqslant\;}
\renewcommand{\geq}{\;\geqslant\;}
\newcommand{\isdefby}{\; := \;}

\newcommand{\dd}{{\rm d}}
\newcommand{\e}[1]{\,{\rm e}^{#1}\,}

\newcommand{\sumtwo}[2]{\sum_{\substack{#1 \\ #2}}}
\newcommand{\sumthree}[3]{\sum_{\substack{#1 \\ #2 \\ #3}}}

\newcommand{\inttwo}[2]{\int_{\substack{#1 \\ #2}}}

\DeclareMathOperator*{\union}{\text{\large$\cup$}}
\newcommand{\Biggerint}[1]{\lceil #1 \rceil}

\newcommand{\dprime}{''}

\def\Tr{{\operatorname{Tr\,}}}

\def\Int{{\operatorname{Int\,}}}
\DeclareMathOperator{\supp}{supp}

\DeclareMathOperator{\Supp}{supp}
\def\diam{{\operatorname{diam\,}}}
\def\dist{{\operatorname{dist\,}}}
\def\supp{{\operatorname{supp\,}}}
\def\core{{\operatorname{core\,}}}
\def\soft{{\operatorname{soft\,}}}
\def\loo{{\operatorname{loop\,}}}
\def\sma{{\operatorname{small\,}}}

\def\Re{{\operatorname{Re\,}}}

\def\per{{\text{\rm\,per}}}

\def\sma{{\text{\rm\,small}}}

\def\bra #1{\langle#1 |\,}
\def\ket #1{\,|#1 \rangle}
\newcommand{\braket}[2]{\langle#1 | #2 \rangle}
\newcommand{\expval}[1]{\langle #1 \rangle}

\newcommand{\neighbours}[2]{<\! #1,#2 \! >}
\newcommand{\compl}{{\text{\rm c}}}
\newcommand{\indicator}[1]{\,\mathbb I \, \bigl[ #1 \bigr]}

\newcommand{\trfunc}{\varphi^{\rm T}}
\newcommand{\Trfunc}{\Phi^{\rm T}}
\newcommand{\ad}{\text{ad}}
\newcommand{\normV}{{\|\bsV\|}}

\newcommand{\VA}{\bsV_{\!\!\! \bsA}}
\newcommand{\VAone}{\bsV_{\!\!\! \bsA_1}}
\newcommand{\VAtwo}{\bsV_{\!\!\! \bsA_2}}
\newcommand{\VAthree}{\bsV_{\!\!\! \bsA_3}}
\newcommand{\VAfour}{\bsV_{\!\!\! \bsA_4}}
\newcommand{\VAi}{\bsV_{\!\!\! \bsA_i}}
\newcommand{\VAj}{\bsV_{\!\!\! \bsA_j}}
\newcommand{\VAk}{\bsV_{\!\!\! \bsA_k}}
\newcommand{\VAell}{\bsV_{\!\!\! \bsA_\ell}}
\newcommand{\VAm}{\bsV_{\!\!\! \bsA_m}}

\def\writefig#1 #2 #3 {\rlap{\kern #1 truecm \raise #2 truecm
\hbox{#3}}}

\newcommand{\caA}{{\mathcal A}}
\newcommand{\caB}{{\mathcal B}}
\newcommand{\caC}{{\mathcal C}}
\newcommand{\caD}{{\mathcal D}}
\newcommand{\caE}{{\mathcal E}}
\newcommand{\caF}{{\mathcal F}}
\newcommand{\caG}{{\mathcal G}}
\newcommand{\caH}{{\mathcal H}}
\newcommand{\caI}{{\mathcal I}}

\newcommand{\caL}{{\mathcal L}}
\newcommand{\caM}{{\mathcal M}}

\newcommand{\caP}{{\mathcal P}}
\newcommand{\caQ}{{\mathcal Q}}
\newcommand{\caR}{{\mathcal R}}
\newcommand{\caS}{{\mathcal S}}
\newcommand{\caT}{{\mathcal T}}
\newcommand{\caU}{{\mathcal U}}

\newcommand{\caW}{{\mathcal W}}

\newcommand{\caY}{{\mathcal Y}}

\newcommand{\bbC}{{\mathbb C}}

\newcommand{\bbL}{{\mathbb L}}

\newcommand{\bbN}{{\mathbb N}}

\newcommand{\bbR}{{\mathbb R}}

\newcommand{\bbT}{{\mathbb T}}

\newcommand{\bbZ}{{\mathbb Z}}

\newcommand{\frz}{{\mathfrak z}}

\newcommand{\frM}{{\mathfrak M}}

\newcommand{\bsm}{{\boldsymbol m}}
\newcommand{\bsn}{{\boldsymbol n}}

\newcommand{\bsA}{{\boldsymbol A}}
\newcommand{\bsB}{{\boldsymbol B}}
\newcommand{\bsC}{{\boldsymbol C}}

\newcommand{\bsH}{{\boldsymbol H}}

\newcommand{\bsV}{{\boldsymbol V}}

\newcommand{\bsomega}{{\boldsymbol \omega}}

\begin{document}

\begin{quote}
\raggedleft
{\small
\version
}
\end{quote}
\vspace{2mm}

\title[Effective interactions due to quantum fluctuations]{Effective
interactions due to quantum fluctuations}

\maketitle
\thispagestyle{empty}

\vspace{1mm}

\renewcommand{\thefootnote}{\roman{footnote}}

\begin{centering}
Roman Koteck\'y$^1$\footnote{Also at Katedra Teoretick\'e Fyziky;  partly
supported by the grants GA\v CR 202/96/0731 and GAUK 96/272.} and Daniel
Ueltschi$^2$\\

\end{centering}

\vspace{2mm}

\begin{centering}
{\small\it
$^1$Centrum pro Teoretick\'a Studia, Universita Karlova, Praha\\
$^2$Institut de Physique Th\'eorique, EPF Lausanne\\
}
\end{centering}

\vspace{5mm}

{\small
\noindent{\bf Abstract.}
Quantum lattice systems are rigorously studied at low
temperatures. When the Hamiltonian of the system consists of a potential
(diagonal) term and a --- small --- off-diagonal matrix containing typically
quantum effects, such as a hopping matrix, we show that the latter creates an
effective interaction between the particles.

In the case that the potential matrix has infinitely many degenerate ground states,
some of them may be stabilized by the effective potential. The low temperature
phase diagram is thus a small deformation of the zero temperature phase diagram
of the diagonal potential {\it and} the effective potential. As illustrations we
discuss the asymmetric Hubbard model and the hard-core Bose-Hubbard model.

\vspace{1mm}

\vspace{3mm}

\tableofcontents
}  

\vfill
\newpage

{\footnotesize
{\sc Roman Koteck\'y \hfill\newline
Center for Theoretical Study, Charles University,\hfill\newline
Jilsk\'a 1, 110 00 Praha 1, Czech Republic
\hfill\newline
\phantom{18.}and \hfill\newline
Department of Theoretical Physics, Charles University,\hfill\newline
V~Hole\v sovi\v ck\'ach~2, 180~00~Praha~8, Czech Republic}

\email{kotecky@cucc.ruk.cuni.cz}

\email{http://www.cts.cuni.cz/\~{}kotecky/}

\vspace{5mm}

{\sc Daniel Ueltschi \hfill\newline
Institut de Physique Th\'eorique,\hfill\newline
Ecole Polytechnique F\'ed\'erale,\hfill\newline
CH--1015 Lausanne, Switzerland}
~~
\email{ueltschi@epfl.ch}

\email{http://dpwww.epfl.ch/instituts/ipt/ueltschi/ueltschi.html}
}

\vfill
\thispagestyle{empty}

\newpage

\setcounter{page}{1}
\setcounter{footnote}{0}
\renewcommand{\thefootnote}{\arabic{footnote}}

\section{Introduction}

Physics of a large number of quantum particles at equilibrium is 
 very interesting and difficult at the same
time. Interesting, because it is treating such
macroscopic phenomena as magnetization, crystallisation, superfluidity or
superconductivity. And difficult, because their study has to combine Quantum Mechanics
and Statistical Physics.

A natural approach is to decrease difficulties arising  from this combination by starting
from only one aspect. Thus one  can use only Quantum Mechanics and treat the particles
 first  as independent, trying next to add small interactions. In the present paper we are
concerned with the other approach. Namely,  to start with a model treated by Classical
Statistical Physics, adding next a small quantum perturbation. Another simplification is
to consider lattice systems (going back to a physical justification for the modeling
process, we can invoke applications to condensed matter physics).

Quantum systems studied here have Hamiltonians consisting of two terms. The first
term is a classical interaction between  particles; formally, this operator is
``function" of the position operators of the particles and it is diagonal with
respect to the corresponding basis in occupation numbers. The
second term is an off-diagonal operator that we suppose to be small with respect to
the interaction. Typical example for this is a hopping matrix.

The aim of the paper is to show that a {\it new effective interaction} appears
that is due to the combination of the potential and the kinetic term. An
explicit formula is computed, and sufficient conditions are given in order that
the  low temperature behaviour is  controlled by the sum of the original
diagonal interaction and the effective potential. To be more precise, it is rigorously shown
that the phase diagram of the original quantum model is only a small perturbation of the
phase diagram of a classical lattice model with the effective interaction.

Thus, we will start by recalling some standard  ideas of Classical Statistical Mechanics of
lattice systems. The Peierls argument for proving the occurrence of a first order phase
transition in the Ising model \cite{Pei,Dob,Gri} marks the beginning of the perturbative
studies of the low temperature regimes of classical statistical models.
Partition functions and expectation values of observables may be expanded with
respect to the  excitations on top of  the ground states, interpreting the excitations in
geometric  terms as  {\it contours}.
These ideas and methods are referred to as the Pirogov-Sinai theory; they were first
introduced in \cite{PS, Sin} and later further extended   \cite{Zah, BI, BS}.

The intuitive picture is that a low temperature phase is essentially a ground
state configuration with small excitations. A phase is stable whenever it is
unprobable to install a large domain with another phase inside. For such an
insertion one has to pay on its boundary,  it is excited (two phases are
separated by excitations), but, on the other side, one may gain on its volume if its {\it
metastable free energy} (its ground energy minus the contribution of small thermal
fluctuations) is
smaller than the one of the external phase. It is important to take into
account the fluctuations since they can play a role in determining which phase is
dominant. A standard example here is the Blume-Capel model with an external field
slightly favouring the ``+1" phase; at low temperatures, the ``0" phase may be still
selected because it has more low energy excitations (theory of such dominant states
chosen by thermal fluctuations may be found in \cite{BS}).

The partition function of a quantum system $\Tr \e{-\beta H}$ may be expressed
using Duhamel expansion (or Trotter formula), yielding a classical contour model
in a space with one more (continuous) dimension. If the corresponding classical model
(the diagonal part only) has stable low temperature phases, and if the off-diagonal
terms of the Hamiltonian are small, the contours have low probability of
occurrence and it is possible to extend the Peierls argument to quantum
models \cite{Gin}.  More generally,  one can formulate a ``Quantum
Pirogov-Sinai theory" \cite{BKU, DFF1}, in order to establish that (i) low
temperature phases are very close to ground states of the diagonal interaction
(more precisely: the density matrix $\frac1Z \e{-\beta H}$ is close to the
projection operator $\ket g \bra g$, where $\ket g$ is the ground state of the
diagonal interaction only) and (ii) low temperature phase diagrams are small
deformations of zero temperature phase diagrams of the interactions.

So far we have only discussed the case when the effect of the quantum perturbation is
small, and the features of the phases are due to the classical interaction between the
particles. It may happen, however, that the classical interaction alone is not sufficient to
choose the low temperature behaviour. This is the case in the two models we introduce
now and use later for illustration of our general approach.

\begin{itemize}
\item {\bf The asymmetric Hubbard model.} It describes hopping spin $\frac12$
particles on a lattice $\Lambda \subset \bbZ^\nu$. A basis of its Hilbert space
is indexed by classical configurations $ n \in \{ 0, \uparrow, \downarrow, 2
\}^\Lambda$, and the Hamiltonian
\be
\label{defHubbard}
H = -\sum_{\|x-y\|_2=1} \sum_{\sigma = \uparrow, \downarrow} t_\sigma
c^\dagger_{x\sigma} c_{y\sigma} + U \sum_x n_{x\uparrow} n_{x\downarrow} - \mu
\sum_x (n_{x\uparrow} + n_{x\downarrow})
\end{equation}
(the hopping parameter $t_\sigma$ depends on the spin of the particle). In the
atomic limit $t_\uparrow = t_\downarrow = 0$ the ground states are all the
configurations with exactly one particle at each site. The degeneracy equals
$2^{|\Lambda|}$, which means that it has nonvanishing residual entropy at zero temperature.

\item {\bf The hard-core Bose-Hubbard model.} We consider bosons moving on a
lattice $\Lambda \subset \bbZ^2$. They interact through an infinite on-site
repulsive potential (hard-core), nearest neighbour and next nearest neighbour
repulsive potentials. A basis of its Hilbert space is the set of all
configurations $n \in \{ 0, 1 \}^\Lambda$, and its Hamiltonian:
\be
\label{defBoseHubbard}
H = -t \sum_{\|x-y\|_2=1} a^\dagger_x a_y + U_1 \sum_{\|x-y\|_2=1} n_x n_y + U_2
\sum_{\|x-y\|_2=\sqrt2} n_x n_y - \mu \sum_x n_x .
\end{equation}
For $U_1 \geq 2U_2$, and if $0<\mu<8U_2$, the ground states of the potential
part are those generated by $\bigl( \begin{smallmatrix} 1 & 0 \\ 0 & 0
\end{smallmatrix} \bigr)$, \ie any configuration with each two lines empty, and
the other antiferromagnetic, is a ground state (and similarly in the other direction). The
degeneracy is of the order
$2^{\frac12 |\Lambda|^{\frac12}}$ (if $\Lambda$ is a square), there is no
residual entropy.
\end{itemize}

In these two situations, the smallest quantum fluctuations yield an effective
interaction, and this interaction stabilizes phases displaying long-range order
(there is neither superfluidity nor superconductivity).

In the case where classical and quantum particles are mixed in one model, 
like the Falicov-Kimball model,  a method
using Peierls argument was proposed by Kennedy and Lieb
\cite{KL}; it was extended in \cite{LM} to situations that are not covered by
the present paper, namely to cases of such mixed systems with continuous classical
variables.

Results very similar to ours have already been obtained by Datta, Fern\'andez,
Fr\"ohlich and Rey-Bellet \cite{DFFR}. Their approach is  different, however.
Starting from a Hamiltonian $H(\lambda) = H^{(0)} + \lambda V$, $H^{(0)}$ being
a diagonal operator with infinitely many ground states, and $V$ the quantum
perturbation, the idea is to choose an antisymmetric matrix $S = \lambda
S^{(1)} + \lambda^2 S^{(2)}$ in such a way that the operator 
$
H^{(2)}(\lambda) = \e S H(\lambda) \e{-S}, 
$
expanded with the help of  Lie-Schwinger series,
turns out to be  diagonal, up to terms of
order $\lambda^3$ or higher. If the diagonal part of $H^{(2)}$ has a finite
number of ground states and the excitations cost strictly positive energy,
 it can be shown that the ground states are stable. It is possible to
include higher orders in this perturbation scheme (see \cite{DFFR}).

In fact, our first intention was to study the stability of the results of \cite{BS}
with respect to a quantum perturbation, and we began the present study as a
warm-up and the first simple step towards this goal. This simple step turned
out  however to be rather involved.
Even though, at the end, the paper contains results similar
to that of \cite{DFFR}, we think that the subject is important enough to
justify an alternative approach, and that there are some advantages in an explicit formula
for the effective potential and sufficient conditions for it to control the low temperature
behaviour that may be useful in explicit applications.

The intuitive background of this paper owes much to the work of Bricmont and
Slawny \cite{BS} discussing the situation with infinite degeneracy of ground states, 
where only a finite number of   ground states is dominating as a result of thermal
fluctuations, and to the  paper of Messager and Miracle-Sol\'e \cite{MM} which was useful
to understand the structure of the quantum fluctuations. Having expanded the partition
function $\Tr
\e{-\beta H}$ using Duhamel formula and having defined {\it quantum contours} as 
excitations with respect to a well chosen classical configuration, we identify the
smallest quantum contours (that we call {\it loops}). Given a set of big quantum
contours, we can replace the sum over sets of loops by an {\it effective interaction}
acting on the quantum configurations without loops. This effective interaction is
long-range, but decays exponentially quickly with respect to the distance.
This allows, for a class of models, to have an explicit control on the approximation given
by effective interaction allowing to prove rigorous statements about the behaviour of
original quantum model.

An important model that does not fall into the class of models we can treat, is the
(symmetric) Hubbard model. Take $U=1$ and $t_\uparrow = t_\downarrow = t$ in
\eqref{defHubbard}. Computing the effective potential stemming from one transition of a
particle to a neighbouring site and back, we find an antiferromagnetic interaction of
strength
$t^2$. On the other hand, it is possible to make two transitions  as a result of which the
spins of nearest neighbours are interchanged,
$$
\ket{n_x,n_y} = \ket{\downarrow,\uparrow} = - c_{x\downarrow}^\dagger
c_{y\downarrow} c_{y\uparrow}^\dagger c_{x\uparrow} \ket{\uparrow,\downarrow}.
$$
It turns out that this brings the factor $t^2$, which is of the same order as the strength of
the effective interaction. In this case we cannot ensure the stability of the
phases selected by the effective potential --- we would need a stronger effective
interaction. Otherwise the system jumps easily from a
configuration with one particle per site to another such configuration, \ie
from a classical ground state to another classical ground state. We call
 {\it quantum instability} this
property of the system. In the Hubbard model it is
a manifestation of a continuous symmetry of the system, namely the rotation
invariance.

In Section \ref{secassstat} the ideas discussed above are introduced with the precise
definitions. The effective potential is written down in Sections
\ref{secdefpot} (a general formula) and \ref{secdefpotsimple} (a simpler
formula in special cases).  The results of the paper are summarized in Theorems
\ref{thmphases} (a characterization of stable pure phases) and \ref{thmphasediagram}
(the structure of the phase diagram); experts will recognize standard
formulations of Pirogov-Sinai theory. Taking into account that our aim is to describe in
rigorous way the behaviour of a quantum system, some care must be given  to the
introduction of stable phases. We define them with the help of an external field
perturbation of the state constructed with periodic boundary conditions. In Section
\ref{secexamples} we apply the results to our two illustrative examples. The rest of the
paper is devoted to the construction of a contour representation (Section
\ref{secDuhamexp}), the proof of the exponential decay of the weights of the contours
(Section \ref{secbounds}), and, finally, the proofs of our claims with the help of contour
expansions of the expectation values of local observables and the standard Pirogov-Sinai
theory (Section \ref{secexpobs}).

Let us end this introduction by noting that given a model which enters our
setting, it is not a straightforward task to apply our theorems. 
One still has to separate the correct leading orders that determine the behaviour of
effective interaction. This situation has the utmost advantage that it should bring much
more pleasure to users, since the most interesting part of the job remains to be done ---
to get intuition and to understand how the system behaves.

\vspace{2mm}
{\bf Acknowledgments.}
We are thankful to Christian Gruber for discussions. R.K. acknowledges
the Institut de Physique Th\'eorique at EPFL, and D.U. the Center for
Theoretical Study at Charles University for hospitality.

\section{Assumptions and statements}
\label{secassstat}

\subsection{Classical Hamiltonian with quantum perturbation}

Let $\bbZ^\nu$, $\nu\ge 2$, be the hypercubic lattice. We use $|x-y| \isdefby
\|x-y\|_\infty$ to denote the distance between two sites $x,y \in \bbZ^\nu$. $\Omega$ is
the finite state space of the system at site $x=0$, $|\Omega| = S < \infty$.
Our standard setting will be to consider  the system on a finite torus $\Lambda
= (\bbZ/L\bbZ)^\nu$ (\ie a finite hypercube with periodic boundary
conditions).  With a slight abuse of notation we identify $\Lambda$ with a
subset of $\bbZ^\nu$ and always assume that it is sufficiently large (to
surpass the range of considered finite range interactions). A {\it classical
configuration} $n_\Lambda$  (occasionally we suppress the index and denote it 
$n$) is an element of $\Omega^\Lambda$. If $A \subset \Lambda$, the restriction
of $n_\Lambda$ to $A$ is also denoted by $n_A$. $\caH_\Lambda$ is the
(finite-dimensional) Hilbert space spanned by the classical configurations, \ie
the set of vectors
$$
\ket v = \sum_{n_\Lambda} a_{n_\Lambda} \ket{n_\Lambda} , \hspace{3mm}
a_{n_\Lambda} \in \bbC ,
$$
with the scalar product
$$
\braket{v}{v'} = \sum_{n_\Lambda} a_{n_\Lambda}^* a_{n_\Lambda}' .
$$
Given two configurations $n_A \in \Omega^A$ and $n_{A'}' \in \Omega^{A'}$, with
$A \cap A' = \emptyset$, it is convenient to define $n_A n_{A'}' \in \Omega^{A
\cup A'}$ to be the configuration coinciding with $n_A$ on $A$ and with
$n_{A'}'$ on $A'$.

The Hamiltonian is a sum of two terms, $\bsH^{\per}_\Lambda =
\bsH_\Lambda^{(0)\per} +
\bsV^{\per}_\Lambda$. The former is the quantum equivalent of a classical interaction,
the latter is the quantum perturbation.

\begin{assumption}{\rm\bf Classical Hamiltonian.}
\label{assdiagpot}

\flushleft
There exists a periodic interaction $ \Phi$ 
(\ie a collection of functions $\Phi_A: \Omega^A \to
\bbR$)  of finite range $R_0$ (\ie
$\Phi_A =0$ whenever $\diam A > R_0$) and period $\ell_0$ such that
$$
\bsH_\Lambda^{(0)\per} \ket{n_\Lambda} = H_\Lambda^{(0)\per}(n_\Lambda)
\ket{n_\Lambda} =
\sum_{A \subset \Lambda} \Phi_A(n_A) \ket{n_\Lambda} ;
$$
for any torus $\Lambda \subset \bbZ^\nu$ of side $L$ that is a multiple of $\ell_0$
 and any
$n_\Lambda
\in
\Omega^\Lambda$.
\end{assumption}

When stressing the dependence on the interaction $\Phi$, we will also use the notation
$H_\Lambda^{(0)\per}(n_\Lambda)=H_\Lambda^{\Phi\per}(n_\Lambda)$.

 Let us suppose
that a fixed collection of reference configurations
 $G\subset\Omega^{\bbZ^\nu}$ is given%
\footnote{In some situations $G$ is simply the set of all ground configurations of
$\Phi$. When
discussing the full phase diagram, however,  we will typically extend the interaction
$\Phi$ to a class of interactions by adding certain ``external fields'' .
The set $G$ then will actually play the role of ground states of the interaction with
a particular values of external fields (the point of maximal coexistence of ground
state phase diagram).}. 
For any $n\in \Omega^{\bbZ^\nu}$, $x\in \bbZ^\nu$,  and any (finite range, but not
necessarily translation invariant) interaction $\Psi$, we use $e_x^\Psi(n)$ to denote the
local contribution to the ``energy'' $\Psi$ of the configuration $n$ at the site $x$,
$e_x^\Psi(n)=\sum_{A\ni x}\frac{\Psi_A(n_A)}{|A|}$.
Notice that $H_\Lambda^{\Psi\per}(n_\Lambda)=\sum_{x\in\Lambda}e_x^\Psi(n)$
for every $n\in \Omega^{\Lambda}$ and $\Lambda$ sufficiently large.
Finally, let $U(x)=\{y\in\bbZ^\nu; |y-x|\leq R_0\}$, $\bar A = \cup_{x \in A}
U(x)$ and $G_A=\{g_A; g\in G\}$,
$A\subset\bbZ^\nu$.

We assume that the local energy gap of excitations is uniformly bounded from below,
while the spread of local energies of reference states is not too big:

\begin{assumption}{\rm\bf Energy gap for classical excitations.}
\label{assgap}

\flushleft
There exist  strictly positive constants $\Delta_0$ and $\delta_0$ 
such that:
\begin{itemize}
\item  For any $x \in \bbZ^\nu$ and any $n_{U(x)} \notin G_{U(x)}$, one has the lower
bound
\be 
e_x^\Phi(n) - \max_{g \in G} e_x^\Phi(g) \geq \Delta_0 ,
\end{equation}
\item and,
\be
\max_{g, g' \in G} \bigl| e_x^\Phi(g) - e_x^\Phi(g')
\bigr| \leq \delta_0.
\end{equation}
\end{itemize}
Furthermore, we assume the following extension property on the set of reference states
$G$: if, for a  connected $A\subset \bbZ^\nu$,  a configuration $n$ is such that $n_{U(x)} \in
G_{U(x)}$ for any
$x\in A$,  then $n_{\bar A} \in G_{\bar A}$.
\end{assumption}

For later purpose, we note the following consequence of this assumption.

\begin{property}
Let $\Phi$ satisfy Assumption \ref{assgap}, $R$ be such that  $R^\nu \leq
\Delta_0/\delta_0$, and $A
\subset\bbZ^\nu$ with $\diam \bar A \leq R$.
Then  any pair of configurations  $g_{\bar
A} \in G_{\bar A}$ and $n_{\bar A} \notin G_{\bar A}$, with $n_{\bar A \setminus A} =
g_{\bar A \setminus A}$, satisfies the lower bound
\be
\label{boundphi}
\sum_{A' \subset \bar A} \Bigl[ \Phi_{A'}(n_{A'}) - \Phi_{A'}(g_{A'}) \Bigr]
\geq R^{-\nu} \Delta_0 .
\end{equation}
\end{property}

{\footnotesize
\begin{proof}
We choose $g' \in G$ such that $g_{\bar A}' = g_{\bar A}$. Then we have
\be
\sum_{A' \subset \bar A} \Bigl[ \Phi_{A'}(n_{A'}) - \Phi_{A'}(g_{A'}) \Bigr] =
\sum_{x \in \bar A} \Bigl( e_x^\Phi(n_A g_{A^\compl}') - e_x^\Phi(g') \Bigr) .
\end{equation}
Since $n_{\bar A} \notin
G_{\bar A}$, there exists at least one site $x \in A$ such that $n_{U(x)} \notin
G_{U(x)}$. From the assumption, this implies that
$$
\sum_{A' \subset \bar A} \Bigl[ \Phi_{A'}(n_{A'}) - \Phi_{A'}(g_{A'}) \Bigr]
\geq \Delta_0 - \sum_{y \in \bar A, y \neq x} \delta_0 .
$$
Using $|\bar A| \leq R^\nu$, we obtain the property.
\end{proof}
}

The quantum perturbation $\bsV^{\per}_\Lambda$ is supposed to be a periodic
quantum interaction. Namely, $\bsV^{\per}_\Lambda$ is a sum of local
operators $\VA$,
$\bsV^{\per}_\Lambda = \sum_{\bsA} \VA$, where $\VA$ has support
$\supp\bsA = A \subset \Lambda$ and $\bsA$ is, in general, a pair $(A, \alpha)$, where the
index $\alpha$ specifies $\VA$ from a possible finite set of operators
with the same support.  We found it useful to label quantum interactions
$\VA$ not only by the interaction domain $A$, but also, say,  by quantum numbers of
participating creation and annihilation operators. 
Thus, for example, the term $\bsA$ might, in the case of the Hubbard model, be a
pair $(\neighbours xy, \uparrow)$ corresponding to the
operator $\VA = c^\dagger_{x,\uparrow} c_{y,\uparrow}$. 
We refer to $\bsA$ as a {\it quantum transition}.

\begin{assumption}{\rm\bf Quantum Perturbations.}
\label{asslocalint}

\flushleft
The collection of operators $\VA$ is supposed to be
periodic%
\footnote{By taking the least common multiple, we can always suppose the same
periodicity for $\Phi$ and $\bsV$. Moreover, whenever a torus $\Lambda$ is
considered, we suppose that its side is a multiple of $\ell_0$.}, with period
$\ell_0$,  with respect to the translations of
$\supp
\bsA$.  The interactions $\VA$  are assumed to satisfy the following condition, for
fermions or bosons, respectively:
\begin{itemize}
\item (Fermions) $\VA$ is a finite sum of even monomials in creation and
annihilation operators of fermionic particles at a given site, \ie
$$
\VA = \sumtwo{(x_1, \sigma_1), \dots, (x_k, \sigma_k)}{(y_1, \sigma_1')
\dots, (y_\ell, \sigma_\ell')} c_{x_1,\sigma_1}^\dagger
\dots c_{x_k,\sigma_k}^\dagger c_{y_1,\sigma_1'} \dots c_{y_\ell,\sigma_\ell'}
$$
with $x_i, y_i \in A$ and $\sigma_i$, $\sigma_i'$ are the internal degrees of
freedom, such as spins; $k+\ell$ must be an even number. The creation and
annihilation operators satisfy the anticommutation relations
$$
\{ c^\dagger_{x,\sigma}, c^\dagger_{y,\sigma'} \} = 0, \hspace{5mm} \{
c_{x,\sigma}, c_{y,\sigma'} \} = 0,
\hspace{5mm} \{ c^\dagger_{x,\sigma}, c_{y,\sigma'} \} = \delta_{x,y}
\delta_{\sigma, \sigma'}.
$$
\item (Spins or bosons) The matrix element
$$
\bra{n_\Lambda} \VA \ket{n_\Lambda'}
$$
is zero whenever $n_{\Lambda \setminus A} \neq n_{\Lambda \setminus A}'$ and
otherwise it depends on $n_A$ and $n_A'$ only.
\end{itemize}
In both cases $\bsV$ is supposed to have an {\it exponential decay} with respect to its
support: defining the norm of a (quantum) interaction $\bsV$ by\footnote{This is a norm,
provided the multiplication of an interaction $\bsV$ by a scalar $\lambda$ is
defined to be $(\lambda \bsV)_\bsA = \lambda^{|A|} \VA$.}
\be
\label{defnorm}
\normV = \max_{\bsA, A \subset \mathbb Z^\nu} \Bigl[ \max_{n_A, n_A' \in
\Omega^A}
|\bra{n_A'} \VA \ket{n_A}| \Bigr]^{1/|A|} ,
\end{equation}
we assume that $\normV<\infty$.
\end{assumption}

When stating our theorems, we shall actually suppose  $\normV$ to be sufficiently small.
Notice also that we do not assume that $\bsV$ is of finite range, the exponential decay
suffices.

\subsection{The effective potential}
\label{secdefpot}

It is actually a cumbersome task to write down a compact formula for the effective
potential in the general case. A lot of notation has to be introduced, and one pays for the
generality by the fact that the resulting formul\ae{} look rather obscure; nevertheless, the
logic behind the following definitions and equations appears rather naturally along the
steps in Section \ref{secDuhamexp}. In the next subsection we shall discuss a special case
where the effective interaction is due to at most four transitions resulting in much simpler
and straightforward formul\ae. We would like to stress that for typical concrete models
this is entirely sufficient. The reader might thus skip the present subsection on the first
reading and consider only the simplified situation of the next subsection.

The real meaning of the next definitions  (in particular, \eqref{defcaI}) will appear more
clearly only in  Section \ref{secDuhamexp}, but, in general case, we cannot leave it aside. 
First of all, we assume that a list $\caS$  of sequences of quantum transitions $\bsA$
is given  to represent the leading quantum fluctuations.
The particular choice of $\caS$ depends on properties of the considered model.
Often the obvious choice like ``any sequence of transitions not surpassing a given order'' is
sufficient. In general case, certain  conditions (specified later in Assumption 5)
involving $\caS$ are to be met.
For any $g_A \in G_A$, the effective potential
$\Psi$ is defined to equal
\bm
\label{defeffpot}
\Psi_A(g_A) = - \sum_{n \geq 1} \frac1{n!} \sum_{k_1, \dots, k_n \geq 2}
\sumtwo{(\bsA^1_1, \dots, \bsA^1_{k_1}, \bsA^2_1, \dots, \bsA^n_{k_n}) \in
\caS}{\cup_{i,j} \bar A^i_j = A} \\
\prod_{i=1}^n \biggl\{ \sum_{n_A^{i,1}, \dots, n_A^{i,k_i-1} \notin G_A}
\caI(A^i_1, \dots, A^i_{k_i}; n_A^{i,1}g_{\Lambda\setminus A}, \dots,
n_A^{i,k_i-1}g_{\Lambda\setminus A}) \Bigl[ \prod_{j=1}^{k_i} \bra{n_A^{i,j-1}}
\bsV_{\!\!\! \bsA^i_j} \ket{n_A^{i,j}} \Bigr] \\
\int_{-\infty < \tau^i_1 < ... < \tau^i_{k_i} < \infty} \dd\tau^i_1 \dots
\dd\tau^i_{k_i} \Bigl[ \prod_{j=1}^{k_i-1} \e{-(\tau^i_{j+1}-\tau^i_j) \sum_{A'
\subset A^i} [\Phi_{A'}(n_{A'}^{i,j}) - \Phi_{A'}(g_{A'})]} \Bigr] \biggr\} \\
\frac{\indicator{\min_{i,j} \tau^i_j < 0 \text{ and } \max_{i,j} \tau^i_j >
0}}{\max_{i,j} \tau^i_j - \min_{i,j} \tau^i_j} \trfunc(B_1, \dots B_n).
\end{multline}
To begin to decode this formula, notice first that  the second sum is over all
sequences\newline
$(\bsA^1_1,
\dots,
\bsA^1_{k_1}, 
\bsA^2_1, \dots, \bsA^n_{k_n})$ of transitions that  are in the list $\caS$
and are just covering  the set $A$, $\cup_{i,j} \bar A^i_j = A$.
The sum in the braces (for a given $i=1,\dots, n$) is taken over collections of configurations
$n_A^{i,1}, \dots, n_A^{i,k_i-1} \notin G_A$ with  $n_A^{i,0} \equiv
n_A^{i,k_i} \equiv g_A$, while the integral is taken over ``times'' attributed to transitions,
with the energy term in the exponent taken over the set $A^i =\union_{j=1}^{k_i} \bar
A^i_j$,
$\bar A = \cup_{x \in A} U(x)$.

Finally, there are some restrictions on the sums and integrals encoded in functions\newline 
$\frac{\indicator{\min_{i,j} \tau^i_j < 0 \text{ and } \max_{i,j} \tau^i_j >
0}}{\max_{i,j} \tau^i_j - \min_{i,j} \tau^i_j} $,  $\trfunc(B_1, \dots B_n)$, and
$\caI(A^i_1, \dots, A^i_{k_i}; n_A^{i,1}g_{\Lambda\setminus A}, \dots,
n_A^{i,k_i-1}g_{\Lambda\setminus A})$.
The easiest  is the first one. One just assumes that the interval between the first and the
last of concerned ``times'' contains the origin and the integrand is divided by the length of
this interval. The function $\trfunc(B_1, \dots B_n)$ in terms of the sets $B_i = A^i \times
[\tau^i_1, \tau^i_{k_i}]\subset \bbZ^\nu \times [-\infty,\infty]$, $i=1,\dots,n$, is the
standard factor from the theory of cluster expansions defined as 
$$
\trfunc(B_1, \dots, B_n) = \begin{cases} 1 & \text{if $n=1$} \\ \sum_{\mathcal
G} \prod_{e(i,j) \in \mathcal G} \bigl( -\indicator{B_i \cup B_j \text{ is connected }} \bigr)
&
\text{if
$n
\geq 2$} \end{cases}
$$
with the sum  over all connected graphs $\mathcal G$ of $n$ vertices.
Connectedness of a set $B \subset \bbZ^\nu \times [-\infty,\infty]$ is
defined by combining connection in continuous direction with connection in slices
$\{x | (x,\tau)\in B\}\subset \bbZ^\nu$ through pairs of sites of distance one.
The most difficult to define is the restriction given by the function $\caI$
that characterizes whether the collection of transitions is connected, in some generalized
sense, through the intertwining configurations. A consolation might be that in lowest orders
it is always true. Namely, 
 $\caI(A_1, \dots, A_k; n^1_\Lambda, \dots,
n^{k-1}_\Lambda)=1$ whenever 
$k\leq5$. To define it in  a general case, consider
$A_1, \dots,
A_k
\subset
\bbZ^\nu$ and
$n^1,
\dots, n^{k-1}
\in \Omega^{\bbZ^\nu}$. Taking $\bar A = \cup_{x \in A} U(x)$ and $E(n) = \{ x
\in \Lambda : n_{U(x)} \neq g_{U(x)} \text{ for any } g \in G \}$, we consider
the set $\hat B^{(0)} \subset \bbZ^{\nu+1}$,
$$
\hat B^{(0)} = \union_{j=1}^k \Bigl[ \bar A_j \times \{2j-2\} \Bigr] \union
\union_{j=1}^{k-1} \Bigl[ E(n^j) \times \{2j-1\} \Bigr] .
$$
Think of layers, one on top of another --- configurations on odd levels interspersed with
transitions on even levels.
The set $\hat B^{(0)}$ decomposes into connected components, $\hat B^{(0)} =
\union_{\ell\geq 1} \hat B^{(0)}_\ell$. To any $\hat B^{(0)}_\ell$, define the
box $\tilde B^{(0)}_\ell \subset \bbZ^{\nu+1}$ as the smallest rectangle
containing $\hat B^{(0)}_\ell$. Then let $\hat B^{(1)} = \union_{\ell\geq 1}
\tilde B^{(0)}_\ell$; decompose into connected components $\hat B^{(1)} =
\union_{\ell\geq 1} \hat B^{(1)}_\ell$, and repeat the procedure until no
change occurs any more, \ie until $\hat B^{(m)} = \union_{\ell\geq 1} \tilde
B^{(m)}_\ell$. The function $\caI$ characterizes whether this final set, the result of the
above construction, is connected or not,  
\be
\label{defcaI}
\caI(A_1, \dots, A_k; n^1_\Lambda, \dots, n^{k-1}_\Lambda) = \begin{cases} 1
&\text{if $\hat B^{(m)}$ is connected} \\ 0 &\text{otherwise.} \end{cases}
\end{equation}

\subsection{Quantum fluctuations with less than four transitions}
\label{secdefpotsimple}

The equation \eqref{defeffpot} for the effective potential is hard to handle
in general case. However, in many situations it is enough to consider  only
small sequences of less than four quantum transitions to define it. We rewrite in this
section the explicit formul\ae{} for the effective potential in such a case.

We assume thus that a list $\caS$  of sequences  of quantum transitions $\bsA$,
containing at most 4 transitions, 
is given  to represent the most important quantum fluctuations.
Let us decompose $\caS = \caS^{(2)}\cup \caS^{(3)}\cup \caS^{(4)}$, with
$\caS^{(k)}$ denoting the list of sequences with exactly $k$ transitions, and write
\be
\Psi = \Psi^{(2)} + \Psi^{(3)} + \Psi^{(4)}.
\end{equation}
Here $\Psi^{(k)}$ is the contribution to the effective potential  due to the fluctuations from
$\caS^{(k)}$.

Let
$$
\phi_A(n_A;g_A) = \sum_{A' \subset A} \Bigl[ \Phi_{A'}(n_{A'}) -
\Phi_{A'}(g_{A'}) \Bigr] .
$$

Then, for any connected $A \subset \bbZ^\nu$ and $g_A \in G_A$, we define
\ba
\label{defPsi2}
\Psi_A^{(2)}(g_A) &= - \sumtwo{(\bsA_1,\bsA_2) \in \caS^{(2)}}{\bar A_1 \cup
\bar A_2 = A} \sum_{n_A \notin G_A} \frac{\bra{g_A} \VAone \ket{n_A}
\bra{n_A} \VAtwo \ket{g_A}}{\phi_A(n_A;g_A)} ,\\
\Psi_A^{(3)}(g_A) &= - \sumtwo{(\bsA_1, \bsA_2, \bsA_3) \in \caS^{(3)}}{ \bar
A_1 \cup \bar A_2 \cup \bar A_3 = A} \sum_{n_A,n_A' \notin G_A} \frac{\bra{g_A}
\VAone \ket{n_A} \bra{n_A} \VAtwo \ket{n_A'} \bra{n_A'}
\VAthree \ket{g_A}}{\phi_A(n_A;g_A) \phi_A(n_A';g_A)} .
\end{align}
The expression for $\Psi^{(4)}$ becomes more complicated (we shall see in
Section \ref{secDuhamexp} that clusters of excitations are actually occurring here),
\bm
\Psi_A^{(4)}(g_A) = - \sumtwo{(\bsA_1, \bsA_2, \bsA_3, \bsA_4) \in
\caS^{(4)}}{\bar A_1 \cup \bar A_2 \cup \bar A_3 \cup \bar A_4 = A} \biggl[
\sum_{n_A,n_A',n_A\dprime \notin G_A} \tfrac{\bra{g_A} \VAone \ket{n_A}
\bra{n_A} \VAtwo \ket{n_A'} \bra{n_A'} \VAthree \ket{n_A\dprime}
\bra{n_A\dprime} \VAfour \ket{g_A}} {\phi_A(n_A;g_A) \phi_A(n_A';g_A)
\phi_A(n_A\dprime;g_A) } \\
- \frac12 \sum_{n_A,n_A' \notin G_A} \tfrac{\bra{g_A} \VAone \ket{n_A}
\bra{n_A} \VAtwo \ket{g_A} \bra{g_A} \VAthree \ket{n_A'} \bra{n_A'}
\VAfour \ket{g_A}}{\phi_{A^1}(n_A;g_A) + \phi_{A^2}(n_A';g_A)} \Bigl\{
\tfrac1{\phi_{A^1}(n_A;g_A)} + \tfrac1{\phi_{A^2}(n_A';g_A)} \Bigr\}^2 \biggr] .
\end{multline}
Above we denoted $A^1 = \bar A_1 \cup \bar A_2$ and $A^2 = \bar A_3 \cup \bar
A_4$. Property \eqref{boundphi} implies that all the denominators are
strictly positive.

These equations simplify further if $\VA$ is a monomial in creation and
annihilation operators; indeed in the sums over intermediate configurations
only one element has to be taken into account.

Notice, finally,  that the diagonal terms in $\bsV$ are not playing any role in the previous
definitions; we consider that they are small, since otherwise we would have
included them into the diagonal potential.

\subsection{Stability of the dominant states}
\label{secstab}
The aim of rewriting a class of quantum transitions  in terms of
the effective potential was  to get a control  over stable low temperature phases.
To this end, the three conditions, expressed first only vaguely and then in precise terms
in the following  Assumptions  4, 5, and 6, must be met. Namely, we suppose that
\begin{itemize}
\item the Hamiltonian corresponding to the sum $\Phi+\Psi$ of the classical (diagonal)
and  effective interactions has a finite number of ground configurations, and its
excitations have strictly positive energy%
\footnote{Again, when exploring a region of phase diagram at once, we have a fixed
finite set of reference configurations that, strictly speaking, turn out to be ground
configurations of the corresponding Hamiltonian for a particular value of ``external
fields''. See below for a more detailed formulation.};
\item the list $\caS$ contains all the lowest quantum fluctuations;
\item there is no ``quantum instability"; the transition probability from a ``ground
state" $g$ to another ``ground state" $g'$ is small compared to the energy cost
of the excitations.
\end{itemize}

Each component of the effective interaction $\Psi_A$ is a mapping $G_A \to
\bbR$; let us first extend it to $\Omega^A \to \bbR$ by putting
$\Psi_A(n_A) = 0$ if $n_A \notin G_A$. 
To give a precise meaning to the first condition, we suppose that a finite number of 
periodic reference configurations $D\subset G$ is given such that the interaction
$\Phi+\Psi$ satisfies the Peierls condition with respect to $D$. We choose a formulation
in which it is very easy to verify the condition and, in addition, it takes into account the
fact that the configurations from $D$ are not necessarily translation invariant.
Namely, we will formulate the condition in terms of a potential $\Upsilon$
that is equivalent to  $\Phi+\Psi$ and is chosen in a suitable way.
Of course, in many particular cases this is not necessary and the condition as stated
below is valid directly  for $\Phi+\Psi$. However, in several important cases treated in
Section 3, the interaction $\Phi+\Psi$ turns out not to be so called $m$-potential and
the use of the equivalent $m$-potential $\Upsilon$ not only simplifies the formulation
of the Peierls condition, but also makes the task of its verification much easier.

We will consider the interactions $\varphi$ and $\phi$ to be
{\it equivalent}%
\footnote{Usual notion of (physically) equivalent interactions (see \cite{Geo}, 
\cite{EFS}) is slightly weaker, but we will not need it here.}
 if, for any finite torus $\Lambda$ and any configuration $n\in\Omega^\Lambda$, one has
$$
H_\Lambda^{\varphi \per}(n)=H_\Lambda^{\phi \per}(n).
$$
Notice that the above  amounts also to the equality
 $$
\sum_{x\in \Lambda}e_x^{\varphi}(n)=\sum_{x\in \Lambda}e_x^{\phi}(n).
$$

\begin{assumption}{\rm\bf Peierls condition.}
\label{assPeiprop}

\flushleft
There exist a finite set of periodic configurations $D\subset G$ with the smallest
common period $L_0$, a constant
$\Delta$ such that 
$\Delta > \normV^k$ for some finite constant $k$,  and  a periodic interaction
$\Upsilon =\{\Upsilon_A \}$ (with period $\ell_0$) that is equi\-va\-lent to
$\Phi+\Psi$ such that the following conditions are satisfied.
The interaction $\Upsilon$ is of a finite range%
\footnote{We will suppose, taking larger $R$ if necessary, that it  is larger or equal to the
range $R_0$ of $\Phi$, as well as to the range of the effective interaction $\Psi$ and to
$L_0$.}
 $R \in \bbN$  such that
\be
R^{\nu} \leq \Delta_0 / \delta_0 ,
\end{equation}
with the constants $\delta_0$ and $\Delta_0$ determined by the interaction $\Phi$ in 
Assumption \ref{assgap}.
The value $e_x^\Upsilon(d)$ is supposed to be translation invariant with respect to $x$ for
any $d\in D$, and the interaction $\Upsilon $  satisfies the following
conditions:
\begin{itemize}
\item
For any $x\in \Lambda$ and any 
$n$ with $n_{U(x)} \notin G_{U(x)}$, one has
$$
e_x^\Upsilon(n) - \max_{g \in G} e_x^\Upsilon(g)  \geq \tfrac12\Delta_0 .
$$
\item
For any $x\in \Lambda$ and any 
$n$ with  $n_{V(x)} \notin D_{V(x)}$, $V(x)=\{y\in\Lambda; |y-x|\le R\}$,  one has
$$
e_x^\Upsilon(n) - \min_{d \in D} e_x^\Upsilon(d) \geq \Delta .
$$
\end{itemize}
\end{assumption}

The following assumption is a  condition demanding that the list $\caS$ should contain
all  transitions that are relevant for the effective potential.
We define
\be
\bsm(\VAone, \dots, \VAk) = \max_{g \in G} \max_{n^1, \dots,
n^{k-1} \notin G} |\bra g \VAone \ket{n^1} \bra{n^1} \VAtwo
\ket{n^2} \dots \bra{n^{k-1}} \VAk \ket g| .
\end{equation}

\begin{assumption}{\rm\bf Completeness of the set of quantum transitions.}
\label{asscompl}

\flushleft
There exists  a  function $b_1(\boldsymbol\cdot)$  with $\lim_{\lambda\to
0}b_1(\lambda)=0$ such that for  any sequence
$(\bsA_1, \dots, \bsA_m) \notin \caS$ with connected $\cup_{i=1}^m \bar A_i$ one has
$$
 \bsm(\VAone, \dots, \bsV_{\!\!\! \bsA_{k_1}})
\bsm(\bsV_{\!\!\! \bsA_{k_1+1}}, \dots, \bsV_{\!\!\! \bsA_{k_2}}) \dots
\bsm(\bsV_{\!\!\! \bsA_{k_{n-1}+1}}, \dots, \VAm) \leq b_1(\normV)\Delta.
$$
\end{assumption}

Finally, we have a  condition assuring that there is {\it no quantum
instability}.
\begin{assumption}{\rm\bf Absence of quantum instability.}
\label{assnoquinst}

\flushleft
There exists  a  function $b_2(\boldsymbol\cdot)$  with $\lim_{\lambda\to
0}b_2(\lambda)=0$ such that
for any
sequence $(\bsA_1, \dots, \bsA_m)$, and  any $g, g' \in G$, $g \neq g'$, one has
$$
\Bigl| \bra{g} \VAone \dots \VAm \ket{g'} \Bigr|
\leq b_2(\normV) \Delta .
$$
\end{assumption}

\subsection{Characterization of stable phases}

Notice first that the specific energy per lattice site of the  configuration
$d\in D$, defined by
\be
\label{approxgren}
e(d )= \lim_{\Lambda \nearrow \bbZ^\nu} \frac1{|\Lambda|} \sum_{A \subset
\Lambda} [\Phi_A(d_A) + \Psi_A(d_A)],
\end{equation}
is equal, according to Assumption 4, to $e_x^\Upsilon(d)$ (whose value does
not depend on $x$).

Our first result concerns the existence of the thermodynamic limit for the
state under periodic boundary conditions. Taking $L_0$ to be the smallest
common period of periodic configurations from $D$, we always consider in the
following the limit over tori $\Lambda\nearrow \bbZ^\nu$ whose sides are
multiples of $L_0$ and $\ell_0$.

\begin{theorem}{\rm\bf Thermodynamic limit.}
\label{thmlimit}

\flushleft
Suppose that the Assumptions \ref{assdiagpot}--\ref{assnoquinst}
are satisfied.   There exist constants
$\varepsilon_0 > 0$ and
$\beta_0 =\beta_0(\Delta)$ (depending on $\nu, S, R, \ell_0$) such that  the limit
\be
\label{perstav}
\expval T_\beta^{\per} =\lim_{\Lambda\nearrow \bbZ^\nu} \frac{\Tr T \e{-\beta
\bsH_\Lambda^{\per}}}{\Tr
\e{-\beta \bsH_\Lambda^{\per}}} 
\end{equation}
exists whenever $\normV \leq \varepsilon_0$, $\beta \geq \beta_0$, 
and $T$ is a local observable.
\end{theorem}

Notice the logic of constants in the theorem above (as well as in the remaining two
theorems stated below). We first choose $\varepsilon_0$.  Then,  for any $\normV \leq
\varepsilon_0$ one can choose $\beta_0$ (depending on $\Delta$ that is determined in
terms of $\bsV$ through the effective potential $\Psi$) such that the claim is valid for the
given $\bsV$ and any $\beta \geq \beta_0(\Delta)$.
With $\normV\to 0$ we may have to go to lower temperatures (higher $\beta$) to keep the
control. Of course, if $\Delta$ does not vanish with vanishing $\normV$ (\ie Assumption 4 is
valid for $\Phi$ alone) as was the case in \cite{BKU}, one can choose the constant $\beta_0$
uniformly in $\normV$.

If there are coexisting phases for a given temperature and Hamiltonian, the
state  $\expval{\boldsymbol\cdot}_\beta^{\per}$ will actually turn out to be a
linear combination of several pure  states. A standard way how to select such a
pure state is to consider a thermodynamic limit with a suitably chosen fixed
boundary condition. In many situations to which the present theory should
apply,  this approach is not easy to implement. The classical part of the
Hamiltonian might actually consist only of on-site terms and to  make the
system ``feel'' the boundary, the truly quantum terms must be used. One
possibility is, of course, to couple the system with the boundary with the help
of the effective potential. The problem here is, however,  that since we are interested in a
genuine quantum model, we would have to introduce the effective potential directly in the
finite volume quantum state.  Expanding this state, in a similar manner as it will be done in
the next section, we would actually obtain a new, boundary dependent  effective
potential. One can imagine that it would be possible to cancel the respective terms by
assuming that the boundary potential satisfies  certain ``renormalizing self-consistency
conditions''. However, the details of such an approach remain to be clarified.

Here we have chosen  another approach. Namely, we construct the pure states by
limits of states $\expval{\boldsymbol\cdot}_\beta^{\Phi^\alpha\per}$, defined by
\eqref{perstav} with 
$\bsH_\Lambda^{\per}=\bsH_\Lambda^{\Phi^\alpha\per}+\bsV_\Lambda^{\per}$, where 
$\Phi^\alpha$ is a  perturbation of the interaction $\Phi$ suitably chosen in such a way
that one approaches the coexistence point from the one phase region.
Consider thus $\caF_{R_0}$, the space of all periodic interactions of range
$R_0$. We say that a state $\expval{\boldsymbol\cdot}_\beta^{\phi\per}$,
$\phi\in\caF_{R_0}$, is {\it unperturbable} if it is insensitive to small perturbations: 
\be
\expval T_\beta^{\phi,\per} =\lim_{\alpha\to 0}\expval
T_\beta^{(\phi+\alpha\psi)\per}
\end{equation}
for every $\psi\in\caF_{R_0}$ and every local observable $T$.
We  define now a state $\expval{\boldsymbol\cdot}_\beta^{*}$ to be a {\it pure stable
state} (with classical potential  $\Phi$ and quantum interaction $\bsV$)   if there
exists a function $(0,\alpha_0)\ni\alpha \to
\Phi^\alpha\in\caF_{R_0}$ so that
$\lim_{\alpha\to 0+}\Phi^\alpha=\Phi$, the states
$\expval{\boldsymbol\cdot}_\beta^{\Phi^\alpha\per}$ are unperturbable, and
\be
\expval T_\beta^{*} =\lim_{\alpha\to 0+}\expval T_\beta^{\Phi^\alpha\per}
\end{equation}
for every local observable $T$.

\begin{theorem}{\rm\bf Pure low temperature phases.}
\label{thmphases}
Under the Assumptions \ref{assdiagpot}--\ref{assnoquinst} and for any $\eta >
0$, there exist $\varepsilon_0 > 0$ and $\beta_0 = \beta_0(\Delta)$ (depending
on $\nu, S, R, \ell_0$) such that if $\normV \leq \varepsilon_0$ and $\beta
\geq \beta_0$, there exists for every $d\in D$ a  function
$f^\beta(d)$ such that  the set $Q=\{d\in D;\, \Re f^\beta(d) = \min_{d' \in D}
\Re f^\beta(d')\}$ characterizes the set of pure stable phases. Namely, for any
$d\in Q$:
\begin{itemize}
\item[a)]  The function $f^\beta(d)$ is equal to the free energy of the system, \ie
$$
f^\beta(d) = -\frac1\beta \lim_{\Lambda \nearrow \bbZ^\nu} \frac1{|\Lambda|}
\log \Tr \e{-\beta \bsH_\Lambda^{\per}} .
$$
\item[b)]  There exists a  pure stable state
$\expval{\boldsymbol\cdot}_\beta^{d}$. Moreover, it is close to the  state
$\ket{d_\Lambda}$ in the sense that for any bounded local observable $T$ and any
sufficiently large $\Lambda$, one has
$$
\Bigl| \expval{T}_\beta^{d} - \bra{d_\Lambda} T \ket{d_\Lambda}
\Bigr| \leq
\eta |\Supp T| \|T\| 
$$
where $\Supp T$ is the support of the operator $T$.
\item[c)] There is an exponential decay of correlations in the state
$\expval{\boldsymbol\cdot}_\beta^{d}$, \ie there exists a constant $\xi^d
> 0$ such that
$$
\Bigl| \expval{T T'}_\beta^{d} - \expval{T}_\beta^{d}
\expval{T'}_\beta^{d} \Bigr| \leq |\Supp T| |\Supp T'| \| T \| \| T' \|
\e{-\dist(\Supp T, \Supp T')/\xi^d}
$$
for any bounded local observables  $T$ and  $T'$.
\item[d)] The state $\expval{\boldsymbol\cdot}_\beta^{\per}$ is a linear
combination of the states $\expval{\boldsymbol\cdot}_\beta^{d}$, $d\in Q$,
with equal weights,
$$
\expval{T}_\beta^{\per}=\frac{1}{|Q|}\sum_{d\in
Q}\expval{T}_\beta^{d}
$$
for each local observable $T$.
\end{itemize}
\end{theorem}

\subsection{Phase diagram}

We now turn  to the phase diagram at low temperatures. Let $r$ be the number of
dominant states, \ie $r = |D|$.  To be able to  investigate the phase diagram, we
suppose that $r-1$  suitable ``external fields'' are added to the Hamiltonian
$\bsH_\Lambda^\per$.   Or, in other words, we suppose that 
classical potential  $\Phi$ and quantum interaction $\bsV$
depend on a vector parameter
$\mu = (\mu_1,\dots,\mu_{r-1})\in \caU$, where $\caU$ is an open set of $\bbR^{r-1}$.
The dependence should be such that the parameters $\mu $ {\it remove the degeneracy}
on the set $D$ of dominant  states. One way how to formulate this condition is to assume a
nonsingularity of the matrix of derivatives 
$\bigl( \frac{\partial e^\mu(d_j)}{\partial\mu_i}  
\bigr)$.

\begin{assumption}
\label{assboundsder}

\flushleft
The potentials $\Phi$ and $\bsV$ are differentiable  with respect to $\mu$ and there exist a
constant $M< \infty$  such that
$$
\sup_{A \subset \bbZ^\nu, n_A \in
\Omega^A}\Bigl| \frac\partial{\partial\mu_i} \Phi_A(n_A) \Bigr|< M
$$
and
$$
\normV +
\sum_{i=1}^{r-1} \Bigl\| \frac{\partial\bsV}{\partial\mu_i} \Bigr\|< M
$$
 for all $\mu \in \caU$.

Further, there exists a point $\mu_0\in \caU$ such that 
$$
e^{\mu_0}(d) = e^{\mu_0}(d') \text{ for all $d, d' \in D$},
$$
and the inverse of the matrix of derivatives
$$
\biggl( \frac\partial{\partial\mu_i} \bigl[ e^{\mu}(d_j) - e^\mu(d_r) \bigr]
\biggr)_{1 \leq i,j \leq r-1}
$$
has a uniform bound for all $\mu \in \caU$.
\end{assumption}

Notice that if for some $d\in D$ one has $e^\mu(d)=e^\mu:= \min_{d'\in D}{e^\mu(d')}$,
then, according to Peierls condition (Assumption 4), the configuration $d$ is actually a
ground state of $\Upsilon$. Thus, the assumption above implies that the zero temperature
phase diagram has a regular structure: there exists a point  $\mu_0 \in \caU$ where all
energies
$e^{\mu_0}(d)$ are equal, $e^{\mu_0}(d)=e^{\mu_0}$, $r$ lines ending in
$\mu_0$ with $r-1$ ground states, $\frac12 r(r-1)$
two-dimensional surfaces whose boundaries are the lines above with $r-2$ ground states,
\dots, $r$ open $(r-1)$-dimensional domains with only one ground state.
Denoting the ($r-|Q|$)-dimensional manifolds corresponding  to the coexistence of a given
set $Q\subset D$ of ground states by
\be
\frM^*(Q) = \Bigl\{ \mu \in \caU ; \Re e^\mu(d) = \min_{d' \in D} \Re e^\mu(d')
\text{ for all $d \in Q$} \Bigr\},
\end{equation}
we can summarize the above structure by saying that the collection 
 $\caP^*= \{ \frM^*(Q) \}_{Q \subset D}$  determines a {\it regular phase diagram}.
Notice, in particular, that $\cup_{Q\subset D}\frM^*(Q) =\caU$,  $\frM^*(Q) \cap \frM^*(Q') =
\emptyset$ whenever
$Q \neq Q'$, while for the closures,  $\overline\frM^*(Q) \cap \overline\frM^*(Q')=
\overline\frM^*(Q\cup Q')$. Here we set $\frM(\emptyset) = \emptyset$.

The statement of the following theorem is that the similar collection 
$\caP= \{ \frM(Q) \}_{Q \subset D}$  of manifolds corresponding to existence of
corresponding stable pure phases for the full model is also a regular phase diagram and
differs only slightly from $\caP^*$. To measure the distance of two manifolds $\frM$ and
$\frM'$, we introduce the Hausdorff distance
$$
\dist_{{\rm H}}(\frM, \frM')= \max(\sup_{\mu\in\frM}\dist (\mu, \frM'),
\sup_{\mu\in\frM'}\dist (\mu, \frM)).
$$

\begin{theorem}{\rm \bf Low temperature phase diagram}
\label{thmphasediagram}

Under the Assumptions \ref{assdiagpot}--\ref{assboundsder} there exist
$\varepsilon_0 > 0$ and $\beta_0 = \beta_0(\Delta)$ such that if\newline
$\normV +
\sum_{i=1}^{r-1} \| \frac\partial{\partial\mu_i} \bsV\| \leq \varepsilon_0$ and
$\beta \geq \beta_0$, there exists a collection of manifolds $\caP^\beta = \{ \frM^\beta(Q)
\}_{Q
\subset D}$ such that
\begin{itemize}
\item [(a)] The collection $\caP^\beta$ determines a regular phase diagram;
\item [(b)] If $\mu \in \frM^\beta(Q)$, the corresponding stable pure state
$\expval\cdot_\beta^d$ exists for every $d\in Q$ and satisfies the properties b), c), and d),
from Theorem
\ref{thmphases};
\item [(c)] The Hausdorff distance $\dist_{{\rm H}}$ between the
manifolds of $\caP^\beta$ and their correspondent in $\caP^*$ is bounded,
$$
\dist_{{\rm H}}(\frM_\lambda^\beta(Q), \frM_\lambda^*(Q)) \leq
O(\e{-\beta} + \normV +
\sum_{i=1}^{r-1} \Bigl\| \frac{\partial\bsV}{\partial\mu_i} \Bigr\|) ,
$$
for all $Q \subset D$.
\end{itemize}
\end{theorem}

The proofs of these theorems are given in the rest of the paper. Expansions of
the partition function and of expectation values of local observables are
constructed, and interpreted as contours of a classical model in one additional
dimension. Then we show that the assumptions for using the standard Pirogov-Sinai theory
are fulfilled, and, with some special care to be taken
due to our definition of stability, the validity of the three theorems follows.

\section{Examples}
\label{secexamples}

\subsection{The asymmetric Hubbard model}

The state space is $\Omega = \{ 0, \uparrow, \downarrow, 2 \}$ and the
Hamiltonian is written in \eqref{defHubbard}. Hence the classical interaction
is
\be
\Phi_{\{ x\}}(n_x) = \begin{cases} 0 & \text{if } n_x = 0 \\ -\mu & \text{if }
n_x = \uparrow \text{ or } n_x = \downarrow \\ U-2\mu & \text{if } n_x = 2
\end{cases}
\end{equation}
($R_0 = 0$). We choose the chemical potential such that $0 < \mu < U$. The set
$G$ is here the set of ground states of $\Phi$, \ie
$$
G = \{ n \in \Omega^{\bbZ^\nu}: n_x = \uparrow \text{ or } n_x = \downarrow
\text{ for any } x \in \bbZ^\nu \} .
$$
Assumption \ref{assgap} holds with $\Delta_0 = \min(\mu, U-\mu)$ and $\delta_0
= 0$.

The quantum perturbation is defined to be
\be
\VA = \begin{cases} t_\uparrow c^\dagger_{x\uparrow} c_{y\uparrow}
&\text{if } \bsA = (\neighbours xy, \uparrow) \\ t_\downarrow
c^\dagger_{x\downarrow} c_{y\downarrow} &\text{if } \bsA = (\neighbours xy,
\downarrow) \end{cases}
\end{equation}
and we always have $A = \{ x,y \}$ for a pair of nearest neighbours $x,y \in
\bbZ^\nu$. $\normV = |t_\uparrow|^{\frac12}$ (if $|t_\uparrow| \geq
|t_\downarrow|$).

The sequence $\caS$ of transitions that we consider is
$$
\caS = \{ (\bsA, \bsA'): \bsA = (\neighbours xy, \uparrow) \text{ and } \bsA' =
(\neighbours yx, \uparrow) \text{ for some } x,y \in \bbZ^\nu, \|x-y\|_2=1 \} .
$$

The effective potential is given by Equation \eqref{defPsi2}. For any $x,y \in
\bbZ^\nu$, nearest neighbours, any configuration $n$ such that $\ket n =
c^\dagger_{x \uparrow} c_{y \uparrow} \ket g$, $g \in G$, has an increase of
energy of
$$
\phi_{\{x,y\}} (n_{\{x,y\}}; g_{\{x,y\}}) = U .
$$
Furthermore we have
\be
\bra{g_{\{x,y\}}} c^\dagger_{x \uparrow} c_{y \uparrow} c^\dagger_{y \uparrow}
c_{x \uparrow} \ket{g_{\{x,y\}}} + \bra{g_{\{x,y\}}} c^\dagger_{y \uparrow}
c_{x \uparrow} c^\dagger_{x \uparrow} c_{y \uparrow} \ket{g_{\{x,y\}}} =
\begin{cases} 1 & \text{if } g_{\{ x,y \}} \in \{ (\uparrow,\downarrow),
(\downarrow,\uparrow) \} \\ 0 &\text{otherwise.} \end{cases}
\end{equation}
Therefore
\be
\Psi_{\{x,y\}} (g_{\{x,y\}}) = \begin{cases} -t_\uparrow^2 /U & \text{if }
g_{\{ x,y \}} \in \{ (\uparrow,\downarrow), (\downarrow,\uparrow) \} \\ 0
&\text{otherwise.} \end{cases}
\end{equation}
Let $V(x) = \{ y
\in \bbZ^\nu: |y-x| \leq 1 \}$. Taking $\Upsilon = \Phi + \Psi$, we have
\be
e_x^\Upsilon(n_{V(x)}) = \Phi_{\{x\}}(n_x) + \tfrac12 \sum_{y: \| x-y \|_2 =
1} \Psi_{\{x,y\}} (n_{\{x,y\}})
\end{equation}
($e_x^\Upsilon(n_{V(x)})$ was initially a function $\Omega^{\bbZ^\nu} \to
\bbR$, but is actually depends on $n_{V(x)}$ only).

The set $D$ has two elements, namely the two chessboard configurations
$d^{(1)}$ and $d^{(2)}$; if $(-1)^x \isdefby \prod_{i=1}^\nu (-1)^{x_i}$,
$$
d^{(1)}_x = \begin{cases} \uparrow & \text{if } (-1)^x = 1 \\ \downarrow &
\text{if } (-1)^x = -1 \end{cases} \hspace{10mm} d^{(2)}_x = \begin{cases}
\uparrow & \text{if } (-1)^x = -1 \\ \downarrow & \text{if } (-1)^x = 1 .
\end{cases}
$$
The last inequality of Assumption \ref{assPeiprop} holds with $\Delta =
\frac12 t_\uparrow^2 / U$.

The maximum of the expression in Assumption \ref{asscompl} is equal to $2U
\max(t_\downarrow^2 / t_\uparrow^2, t_\uparrow^2)$. If there exists
$\varepsilon>0$ such that $|t_\downarrow| \leq |t_\uparrow|^{1+\varepsilon}$,
the bound of Assumption \ref{asscompl} can be chosen to be $b_1 = 2U
|t_\uparrow|^{2\varepsilon}$. For Assumption \ref{assnoquinst} the expression
has maximum equals to $2U|t_\downarrow| / |t_\uparrow|$ and we can take $b_2 =
2U|t_\uparrow|^\varepsilon$ (this Assumption is not true in the symmetric
Hubbard model; the effective potential is not strong enough in order to forbid
the model to jump from one $g$ to another $g'$).

Our results for the asymmetric Hubbard model can be stated in the following
theorem (first obtained by \cite{DFF2}).

\begin{theorem}[Chessboard phases in asymmetric Hubbard model]

Consider the lattice $\bbZ^\nu$, $\nu \geq 2$, and suppose $0 < \mu < U$ and
$|t_\downarrow| \leq |t_\uparrow|^{1+\varepsilon}$ with $\varepsilon>0$. Then
for any $\delta>0$, there exist $t_0>0$ and $\beta_0(t_\uparrow) < \infty$
($\lim_{t_\uparrow \to 0} \beta_0(t_\uparrow) = \infty$) such that if
$|t_\uparrow| \leq t_0$ and $\beta \geq \beta_0$,
\begin{itemize}
\item The free energy exists in the thermodynamic limit with periodic
boundary conditions, as well as expectation values of observables.
\item There are two pure periodic phases, $\expval\cdot_\beta^{(1)}$ and
$\expval\cdot_\beta^{(2)}$, with exponential decay of correlations.
\item One of these pure phases, $\expval\cdot_\beta^{(1)}$, is a small deformation of
the chessboard state $\ket{d^{(1)}}$:
$$
\expval{n_{x \uparrow}}_\beta^{(1)} \begin{cases} \geq 1-\delta & \text{if }
(-1)^x = 1 \\ \leq \delta & \text{if } (-1)^x = -1 \end{cases} \hspace{10mm}
\expval{n_{x \downarrow}}_\beta^{(1)} \begin{cases} \leq \delta & \text{if }
(-1)^x = 1 \\ \geq 1-\delta & \text{if } (-1)^x = -1 . \end{cases}
$$
The other pure phase, $\expval\cdot_\beta^{(2)}$, is a small deformation of
$\ket{d^{(2)}}$.
\end{itemize}
\end{theorem}

To construct the two pure phases, one way is to consider the Hamiltonian
$$
\bsH_\Lambda^\per(h) = \bsH_\Lambda^\per - h \sum_{x \in \Lambda} (-1)^x
(n_{x\uparrow} - n_{x\downarrow}) .
$$
Then
$$
\expval\cdot_\beta^{(1)} = \lim_{h \to 0+} \expval\cdot_\beta^\per(h)
$$
and
$$
\expval\cdot_\beta^{(2)} = \lim_{h \to 0-} \expval\cdot_\beta^\per(h) ,
$$
where $\expval\cdot_\beta^\per(h)$ is defined by \eqref{perstav} with Hamiltonian
$\bsH_\Lambda^\per(h)$.

\subsection{The hard-core Bose-Hubbard model}

The state space is $\Omega = \{0,1\}$ and the Hamiltonian is written in
\eqref{defBoseHubbard}. Let $P$ a plaquette of four sites; the classical
interaction is
\be
\Phi_P(n_P) = \tfrac12 U_1 \sumtwo{x,y \in P}{\| x-y \|_2 =1} n_x n_y + U_2
\sumtwo{x,y \in P}{\| x-y \|_2 = \sqrt2} n_x n_y - \tfrac14 \mu \sum_{x \in P}
n_x ,
\end{equation}
and $\Phi_A=0$ if $A$ is not a plaquette. Here $R_0 = 1$. Remark that we have
\be
\Phi_P(n_P) = (\tfrac14 U_1 - \tfrac12 U_2) \sumtwo{x,y \in P}{\| x-y \|_2 =1}
(n_x + n_y - \tfrac12)^2 + U_2 \Bigl(
\sum_{x \in P} n_x - \tfrac12 - \frac\mu{8U_2} \Bigr)^2 + C
\end{equation}
with $C = -\frac12 U_1 + \frac34 U_2 - \frac18 \mu - \frac1{64} \mu^2/U_2$.
$\Phi_P(n_P)$ is minimum if $n_P = \bigl( \begin{smallmatrix} 1 & 0 \\ 0 & 0
\end{smallmatrix} \bigr)$, or any configuration obtained from $\bigl(
\begin{smallmatrix} 1 & 0 \\ 0 & 0 \end{smallmatrix} \bigr)$ by rotation. Hence
we define
$$
G = \Bigl\{ n \in \{0,1\}^{\bbZ^2} : n_P \in \bigl\{
\bigl( \begin{smallmatrix} 1 & 0 \\ 0 & 0 \end{smallmatrix} \bigr),
\bigl( \begin{smallmatrix} 0 & 1 \\ 0 & 0 \end{smallmatrix} \bigr),
\bigl( \begin{smallmatrix} 0 & 0 \\ 0 & 1 \end{smallmatrix} \bigr),
\bigl( \begin{smallmatrix} 0 & 0 \\ 1 & 0 \end{smallmatrix} \bigr) \bigr\}
\text{ for any plaquette $P$} \Bigr\}
$$
($G$ is here the set of ground states of the interaction $\Phi$). Since
$\Phi_P(n_P) - \Phi_P(g_P) \geq \frac14 \min(\mu, 8U_2 - \mu)$, for any $n_P
\notin G_P$, $g_P \in G_P$, Assumption \ref{assgap} holds with $\Delta_0 =
\frac1{16} \min(\mu, 8U_2 - \mu)$, provided $0<\mu<8U_2$. $\delta_0 = 0$.

We take as sequence of transitions for the smallest quantum fluctuations
$$
\caS = \{ (\bsA,\bsA') : \bsA = \neighbours xy \text{ and } \bsA' = \neighbours
yx \text{ for some } x,y \in \bbZ^2, \|x-y\|_2 = 1 \} .
$$
The effective potential follows from \eqref{defPsi2}. Let $\caP_{xy} = \cup_{P
\cap \{x,y\} \neq \emptyset} P$ and more generally we denote by $\caP$ any
$3\times4$ or $4\times3$ rectangle. Up to rotations, we have to take into
account five configurations, namely
$$
\begin{smallmatrix} \begin{smallmatrix} 0 & 1 & 0 \\ 0 & 0 & 0 \\ 0 & 1 & 0 \\
0 & 0 & 0 \end{smallmatrix} \\ g^{(A)}_\caP \end{smallmatrix} \hspace{8mm}
\begin{smallmatrix} \begin{smallmatrix} 0 & 1 & 0 \\ 0 & 0 & 0 \\ 1 & 0 & 1 \\
0 & 0 & 0 \end{smallmatrix} \\ g^{(B)}_\caP \end{smallmatrix} \hspace{8mm}
\begin{smallmatrix} \begin{smallmatrix} 1 & 0 & 1 \\ 0 & 0 & 0 \\ 0 & 1 & 0 \\
0 & 0 & 0 \end{smallmatrix} \\ g^{(C)}_\caP \end{smallmatrix} \hspace{8mm}
\begin{smallmatrix} \begin{smallmatrix} 1 & 0 & 1 \\ 0 & 0 & 0 \\ 1 & 0 & 1 \\
0 & 0 & 0 \end{smallmatrix} \\ g^{(D)}_\caP \end{smallmatrix} \hspace{8mm}
\begin{smallmatrix} \begin{smallmatrix} 1 & 0 & 0 \\ 0 & 0 & 1 \\ 1 & 0 & 0 \\
0 & 0 & 1 \end{smallmatrix} \\ g^{(E)}_\caP \end{smallmatrix}
$$
We find $\Psi_\caP(g^{(A)}_\caP) = -t^2/2U_1$, $\Psi_\caP(g^{(C)}_\caP) =
-t^2/4U_2$, and $\Psi_\caP(g^{(B)}_\caP) = \Psi_\caP(g^{(D)}_\caP) =
\Psi_\caP(g^{(E)}_\caP) = 0$.

The equivalent potential $\Upsilon$ should have a range bigger or equal to 3;
in this case, the corresponding energies $e^\Upsilon_x(n)$ would depend on the
configuration on the square $7 \times 7$ centered at $x$. We could proceed in
this way, but actually it simplifies a lot to do the following. Looking in the
derivation of the contour model (Section \ref{Duhamexp}), we see that beside of
Assumption \ref{assPeiprop}, the only property that $e^\Upsilon_x$ has to
satisfy is that
$$
\sum_{A \subset \Lambda^\per} \Bigl[ \Phi_A(n_A) + \Psi_A(n_A) \Bigr] = \sum_{x
\in \Lambda^\per} e^\Upsilon_x(n)
$$
[see equation \eqref{replacement}].
Therefore we do not define $\Upsilon$, but we do define
\be
e_x^\Upsilon(n_{V(x)}) = \tfrac14 \sum_{P \ni x} \Phi_P(n_P) + \tfrac14
\sum_{y, \|y-x\|_2 = 1} \Psi_{\caP_{xy}}(n_{\caP_{xy}}) + \tfrac14 \sum_{\caP
\subset V(x)} \Psi_\caP(n_\caP),
\end{equation}
where $V(x)$ is the square $5 \times 5$ centerd at $x$.

The configurations $g_{V(x)} \in G_{V(x)}$ are (up to rotations and
reflections)
$$
\begin{smallmatrix} \begin{smallmatrix} 1 & 0 & 1 & 0 & 1 \\ 0 & 0 & 0 & 0 & 0
\\ 1 & 0 & 1 & 0 & 1 \\ 0 & 0 & 0 & 0 & 0 \\ 1 & 0 & 1 & 0 & 1
\end{smallmatrix} \\ g^{(a)}_{V(x)} \end{smallmatrix} \hspace{5mm}
\begin{smallmatrix} \begin{smallmatrix} 1 & 0 & 1 & 0 & 1 \\ 0 & 0 & 0 & 0 & 0
\\ 1 & 0 & 1 & 0 & 1 \\ 0 & 0 & 0 & 0 & 0 \\ 0 & 1 & 0 & 1 & 0
\end{smallmatrix} \\ g^{(b)}_{V(x)} \end{smallmatrix} \hspace{5mm}
\begin{smallmatrix} \begin{smallmatrix} 1 & 0 & 1 & 0 & 1 \\ 0 & 0 & 0 & 0 & 0
\\ 0 & 1 & 0 & 1 & 0 \\ 0 & 0 & 0 & 0 & 0 \\ 0 & 1 & 0 & 1 & 0
\end{smallmatrix} \\ g^{(c)}_{V(x)} \end{smallmatrix} \hspace{5mm}
\begin{smallmatrix} \begin{smallmatrix} 1 & 0 & 1 & 0 & 1 \\ 0 & 0 & 0 & 0 & 0
\\ 0 & 1 & 0 & 1 & 0 \\ 0 & 0 & 0 & 0 & 0 \\ 1 & 0 & 1 & 0 & 1
\end{smallmatrix} \\ g^{(d)}_{V(x)} \end{smallmatrix} \hspace{5mm}
\begin{smallmatrix} \begin{smallmatrix} 0 & 1 & 0 & 1 & 0 \\ 0 & 0 & 0 & 0 & 0
\\ 1 & 0 & 1 & 0 & 1 \\ 0 & 0 & 0 & 0 & 0 \\ 0 & 1 & 0 & 1 & 0
\end{smallmatrix} \\ g^{(e)}_{V(x)} \end{smallmatrix} \hspace{5mm}
\begin{smallmatrix} \begin{smallmatrix} 0 & 0 & 0 & 0 & 0 \\ 1 & 0 & 1 & 0 & 1
\\ 0 & 0 & 0 & 0 & 0 \\ 1 & 0 & 1 & 0 & 1 \\ 0 & 0 & 0 & 0 & 0
\end{smallmatrix} \\ g^{(f)}_{V(x)} \end{smallmatrix} \hspace{5mm}
\begin{smallmatrix} \begin{smallmatrix} 0 & 0 & 0 & 0 & 0 \\ 1 & 0 & 1 & 0 & 1
\\ 0 & 0 & 0 & 0 & 0 \\ 0 & 1 & 0 & 1 & 0 \\ 0 & 0 & 0 & 0 & 0
\end{smallmatrix} \\ g^{(g)}_{V(x)} \end{smallmatrix} \hspace{5mm}
\begin{smallmatrix} \begin{smallmatrix} 0 & 0 & 0 & 0 & 0 \\ 0 & 1 & 0 & 1 & 0
\\ 0 & 0 & 0 & 0 & 0 \\ 0 & 1 & 0 & 1 & 0 \\ 0 & 0 & 0 & 0 & 0
\end{smallmatrix} \\ g^{(h)}_{V(x)} \end{smallmatrix}
$$

We find that $e_x^\Upsilon(g_{V(x)}^{(d)}) = e_x^\Upsilon(g_{V(x)}^{(e)})
= e_x^\Upsilon(g_{V(x)}^{(g)}) = -t^2/2U_1 - t^2/4U_2$, otherwise
$e_x^\Upsilon(g_{V(x)}) \geq -3t^2/4U_1 - t^2/8U_2$ (if $U_1 \geq 2U_2$).
This allows to take $\Delta = t^2 (\frac1{8U_2} - \frac1{4U_1})$ in Assumption
\ref{assPeiprop}. The set of dominant states $D$ consists in all the
configurations generated by $g_{V(x)}^{(d)}$ and $g_{V(x)}^{(f)}$. $|D| = 8$.

The maximum of the expression in Assumption \ref{asscompl} is $b_1 = t^2
(\frac1{8U_2} - \frac1{4U_1})^{-1}$. In Assumption \ref{assnoquinst} $b_2=0$,
because $g\neq g'$ means that $g$ and $g'$ must differ on a whole row, and the
matrix element is zero for any finite $m$.

These eight dominant states bring eight pure periodic phases,
$\expval\cdot_\beta^{(1)}, \dots, \expval\cdot_\beta^{(8)}$; each one can be constructed by
adding a suitable field in the Hamiltonian (\eg the projector onto the dominant
state).

\begin{theorem}[Hard-core Bose-Hubbard model]
Consider the hard-core Bose-Hubbard model on the lattice $\bbZ^2$, and suppose
$U_1 > 2U_2$ and $0<\mu<8U_2$. There exist $t_0>0$ and $\beta_0(t) < \infty$
($\lim_{t \to 0} \beta_0(t) = \infty$) such that if $t \leq t_0$ and $\beta \geq
\beta_0$,
\begin{itemize}
\item the free energy exists in the thermodynamic limit with periodic
boundary conditions, as well as
expectation values of observables,
\item there are 8 pure periodic phases with exponential decay of correlations.
\end{itemize}
\end{theorem}

Each of these eight phases is a perturbation of a dominant state $d$, and the
expectation value of any operator is close to its value in the state $d$, see
Theorem \ref{thmphases} for more precise statement.

\section{Contour representation of a quantum model}
\label{secDuhamexp}

Our Hamiltonian has periodicity $\ell_0 < \infty$. 
Without loss of generality, however, one can
consider only translation invariant Hamiltonians, applying the standard trick. 
Namely, if $\Omega$ is
the single site phase space, we let $\Omega' =
\Omega^{\{ 1, \dots, \ell_0 \}^\nu}$; $S' = |\Omega'| = S^{\ell_0^\nu}$. Then
we consider the torus $\Lambda' \subset \bbZ^\nu$, $\ell_0^\nu |\Lambda' |= |\Lambda|$,
each point of which is representing a block of sites in $\Lambda$ of size $\ell_0^\nu$,
and identify
$$
{\Omega'}^{\Lambda'} \simeq \Omega^\Lambda .
$$
Constructing $\caH'$ as the Hilbert space spanned by the elements of
${\Omega'}^{\Lambda'}$, it is clear that $\caH'$ is isomorphic to $\caH$. 
The new translation invariant interactions $\Phi'$ and $ \bsV'$ are defined   
by resumming, for each
$A\subset\Lambda'$, the corresponding contributions with supports in the union of
corresponding blocks. Notice the change in range of interactions.
Namely, it decreased to $\Biggerint{R/\ell_0}$ (the lowest integer bigger or equal to
$R/\ell_0$).

From now on, keeping the original notation $\caH$, $S$, \dots, we suppose that the
Hamiltonian is translation invariant.

The partition function of a quantum model is a trace over a Hilbert space. But
expanding $\e{-\beta \bsH}$ with the help of Duhamel formula we can reformulate
it in terms of the partition function of a classical model in a space with one
additional dimension (the extra dimension being continuous). In this section we
present such an expansion, leading to a contour representation, of the
partition function $Z_\Lambda^{\per} \isdefby \Tr\e{-\beta \bsH_\Lambda^{\per}}$ in a
finite torus $\Lambda$.

Expansion with the help of Duhamel formula yields
\bm
\label{Duhamexp}
\e{-\beta \bsH_\Lambda^{\per}} = \sum_{m \geq 0} \sumtwo{\bsA_1, \dots,
\bsA_m}{\bar A_i \subset \Lambda} \int_{0<\tau_1<...<\tau_m<\beta} \dd\tau_1
\dots \dd\tau_m \\
\e{-\tau_1 \bsH_\Lambda^{(0)\per}} \VAone \e{-(\tau_2-\tau_1)
\bsH_\Lambda^{(0)\per}} \VAtwo \dots \VAm \e{-(\beta-\tau_m)
\bsH_\Lambda^{(0)\per}} .
\end{multline}
Inserting the expansion of unity $\bbbone_{\mathcal H_\Lambda} =
\sum_{n_\Lambda} \ket{n_\Lambda}\bra{n_\Lambda}$ to the right of operators
$\VAj$, we obtain
\bm
\label{Zexp}
Z_\Lambda^{\per} = \sum_{m \geq 0} \sum_{n_\Lambda^1, \dots n_\Lambda^m}
\sumtwo{\bsA_1, \dots, \bsA_m}{\bar A_i \subset \Lambda}
\int_{0<\tau_1<...<\tau_m<\beta} \dd\tau_1 \dots \dd\tau_m \\
\e{-\tau_1  H^{(0)\per}_\Lambda(n_\Lambda^1)} \bra{n_\Lambda^1}
\VAone \ket{n_\Lambda^2} \e{-(\tau_2-\tau_1) 
H^{(0)\per}_\Lambda(n_\Lambda^2)} \dots \bra{n_\Lambda^m} \VAm
\ket{n_\Lambda^1} \e{-(\beta-\tau_m)  H^{(0)\per}_\Lambda(n_\Lambda^1)} .
\end{multline}
This expansion can be interpreted as a classical partition function on the
$(\nu+1)$-dimensional space $\Lambda\times[0,\beta]$. Namely, calling the
additional dimension ``time direction", the partition function $Z_\Lambda^{\per}$ is
a (continuous) sum over all space-time configurations $\bsn_\Lambda =
\bsn_\Lambda(\tau)$, $\tau \in [0,\beta]$, and all possible transitions at
times corresponding to discontinuities of $\bsn_\Lambda(\tau)$. Notice that
$\bsn_\Lambda(\tau)$ is periodic in the time  direction. Thus, actually, we
obtain a classical partition function on the $d+1$-dimensional torus
$\bbT_{\Lambda}=\Lambda\times[0,\beta]_{\operatorname{per}}$ with a circle
$[0,\beta]_{\operatorname{per}}$ in time direction (for simplicity we omit in
$\bbT_{\Lambda}$ a reference to  $\beta$). Introducing the {\it quantum
configuration} $\bsomega_{\bbT_{\Lambda}}$ consisting of the space-time
configuration $\bsn_\Lambda(\tau)$ and the transitions $(\bsA_i,\tau_i)$ at
corresponding times, we can rewrite \eqref{Zexp} in a compact form
\be
\label{Zcomp}
Z_\Lambda^{\per} = \int \dd\bsomega_{\bbT_{\Lambda}}
\rho^{\per}(\bsomega_{\bbT_{\Lambda}})
\end{equation}
with $\rho^{\per}(\bsomega_{\bbT_{\Lambda}})$ standing for the second line of
\eqref{Zexp}.

Now, we are going to specify excitations within a spacetime configuration
$\bsn$ and identify  classes of small excitations --- {\it the
loops}\footnote{Even though the present framework is more general, the name 
comes from thinking about simplest excitations in Hubbard type models. Namely,
a jump of an electron to a neighbouring site and returning afterwards to its
original position.} --- and large ones --- {\it the quantum contours}.

A configuration $n \in \Omega^{\bbZ^\nu}$ is said to be in the state $g \in G$
at site $x$ whenever $n_{U(x)} = g_{U(x)}$ (notice that, in general, $g$ is not
unique). If there is no such $g \in G$, the configuration $n$ is said to be
{\it classically excited} at $x$. We use $E(n)$ to denote the set of all
classically excited sites of $n \in \Omega^{\bbZ^\nu}$. For any
$\Lambda\subset\bbZ^\nu$, let us consider the set $\caQ^{\per}_{\Lambda}$ of
quantum configurations on the torus $\bbT_{\Lambda}$. Whenever
$\bsomega\in\caQ^d_{\Lambda}$, its {\it boundary} $\boldsymbol
B^{(0)}(\bsomega)\subset \bbT_{\Lambda}$ is defined as the union
\be
\label{bound}
\bsB^{(0)}(\bsomega)=(\cup_{\tau\in[0,\beta]} ( E(\bsn(\tau)) \times \tau)
)\cup(\cup_{i=1}^m (\bar A_i\times \tau_i)) .
\end{equation}
The sets $\bar A_i\times \tau_i \subset \bbT_{\Lambda}$ represent the effect of
the operator $\bsV$ and for this reason are called {\it quantum transitions}.
It is worth to notice that the set $\boldsymbol B^{(0)}(\bsomega)$ is closed.

Next step is to identify the smallest quantum excitations --- those consisting
of a sequence of transitions from the list $\mathcal S$. First, let us use
$\caB^{(0)}(\bsomega)$ to denote the set of connected components of
$\bsB^{(0)}(\bsomega)$ (so that $\bsB^{(0)}(\bsomega)=\cup_{B\in \mathcal
B^{(0)}(\bsomega)}B$). To any $B\in \caB^{(0)}(\bsomega)$  that is not wrapped
around the cylinder (i.e., for which there exists time
$\tau_B\in[0,\beta]_{\operatorname{per}}$ with $B\cap (\bbZ^\nu\times
\tau_B)=\emptyset$) we assign its sequence of transitions, $S(B, \bsomega)$,
ordered according to their times (starting from $\tau_B$ to $\beta$ and
proceeding from $0$ to $\tau_B$) as well as the smallest box $\tilde B$ 
containing $B$. Here, a box is any subset of $\bbT_{\bbZ^\nu}$ of the form
$A\times[\tau_1,\tau_2]$ with connected $A\subset \bbZ^\nu$ and
$[\tau_1,\tau_2]\subset[0,\beta]_{\operatorname{per}}$ (if $\tau_1>\tau_2$, we
interpret the segment $[\tau_1,\tau_2]$ as that interval in
$[0,\beta]_{\operatorname{per}}$ (with endpoints $\tau_1$ and $\tau_2$) that
contains the point $0\equiv\beta$).

We would like to declare the excitations with $S(B,\bsomega)\in \mathcal S$ to
be small. However, we need to be sure that there are no other excitations in
their close neighbourhood. If this were the case, we would ``glue'' the
neighbouring excitations together. This motivates the following iterative
procedure.

Given $\bsomega$, let us first consider the set $\mathcal B_0^{(0)}(\bsomega)$
of those components $B\in \caB^{(0)}(\bsomega)$ that are not wrapped around the
cylinder and for which $S(B,\bsomega)\in\bar\mathcal S$, where $\bar\mathcal S$
is the set of all subsequences of sequences from  $\mathcal S$. Next, we define
the first extension of the boundary,
$$
\bsB^{(1)}(\bsomega)=(\cup_{B\in \caB^{(0)}(\bsomega) \setminus
\caB_0^{(0)}(\bsomega)}B) \cup(\cup_{B\in \mathcal B_0^{(0)}(\bsomega)}\tilde 
B).
$$
Using  $\mathcal B^{(1)}(\bsomega)$ to denote the set of connected components
of $\bsB^{(1)}(\bsomega)$ and $\mathcal B_0^{(1)}(\bsomega)\subset \mathcal
B^{(1)}(\bsomega)$ the set of those components $B$ in $\mathcal
B^{(1)}(\bsomega)$ that are not wrapped around the cylinder and for which%
\footnote{A set $B\in \mathcal B^{(1)}_0(\bsomega)$ may actually contain
several original components from $\mathcal B^{(0)}_0(\bsomega)$. We take for
$S(B,\bsomega)$ the sequence of all transitions in all those components.}
$S(B,\bsomega)\in\bar \mathcal S$, we define
$$
\bsB^{(2)}(\bsomega)=(\cup_{B\in \caB^{(1)}(\bsomega) \setminus
\caB_0^{(1)}(\bsomega)}B) \cup(\cup_{B\in \mathcal B_0^{(1)}(\bsomega)}\tilde 
B) .
$$

Iterating this procedure, it is clear that after finite number of steps we
obtain the final extension of the boundary,
$$
\bsB(\bsomega)=(\cup_{B\in \caB^{(k)}(\bsomega)\setminus \mathcal
B_0^{(k)}(\bsomega)}B) \cup(\cup_{B\in \mathcal B_0^{(k)}(\bsomega)}B).
$$
Here, every $B\in \mathcal B_0^{(k)}(\bsomega)$ is actually a box of the form
$A\times [\tau_1,\tau_2]$ (that is not wrapped around the cylinder) and
$S(B,\bsomega)\in\bar\caS$. Let us denote $\mathcal
B(\bsomega)\equiv\mathcal B_0^{(k)}(\bsomega)$ and consider  the set 
$\mathcal
B_0(\bsomega)\subset\mathcal B(\bsomega)$ of all those sets
$B\in\mathcal B_0^{(k)}(\bsomega)$ for which actually $S(B,\bsomega)\in\caS$
and,  moreover,   $n_A(\tau_1-0)=n_A(\tau_2+0)$. Finally, let $\caB_l(\bsomega)
= \caB(\bsomega) \setminus \caB_0(\bsomega)$ --- it represents the set of all
excitations of $\bsomega$ that are not loops. Taking, for any closed
$B\subset\bbT_{\Lambda}$, the restriction $\bsn_B$ of a space-time
configuration $\bsn$ to be defined by $(\bsn_B)_x(\tau) = \bsn_x(\tau)$ for any
$x\times \tau \in B$, we  introduce the useful notion of the restriction
$\bsomega_B$ of a quantum configuration $\bsomega$ to  $B $ as to consist of
$\bsn_B$ and those quantum transitions from $\bsomega$ that are contained in
$B$,  $A \times \tau \subset B$ (we suppose here that $\bsomega$ and $B$ are
such that no transition intersects both $B$ and its complement; we do not
define $\bsomega_B$ in this case).

Now the loops and and the quantum contours can be defined. First, the {\it
loops} of a quantum configuration $\bsomega$ are the triplets $\xi\equiv (B,
\bsomega_B,g_A^\xi)$; $B \equiv A\times[\tau_1,\tau_2] \in \caB_0(\bsomega)$ is
the {\it support} of the loop $\xi$ and $g_A^\xi=n_A(\tau_1-0)=n_A(\tau_2+0)$,
a restriction of a configuration $g\in G$. (While the configuration $g$ is not
unique, its restriction to $A$ is determined by the loop $\xi$ in a unique
way.) We say that $\xi$ is {\it immersed} in $g$. Given a quantum configuration
$\bsomega$, we obtain a new configuration $\check\bsomega$ by erasing all loops
$(B, \bsomega_B, g^\xi_A)$, \ie for each $\xi$ we remove all the transitions in
its support $B$ and change the space-time configuration on $B$ into $g \in G$
into which $\xi$ is immersed. Let us remark that $\caB(\check\bsomega) =
\caB_l(\bsomega)$. Notice that, since we started our construction from \eqref{bound},
we have automatically $\diam A \ge 2R_0$ for a support $A\times [\tau_1,\tau_2]$
of any loop $\xi$.

{\it Quantum contours} of a configuration $\bsomega$ will be constructed by
extending  pairs $(B, \bsomega_B)$ with $B \in \caB_l(\bsomega)$ by including
also the regions of nondominating states from $G$. Namely, summing over loops
we will see that ``loop free energy'' favours the regions with dominating
configurations from $D\subset G$.  However, to recognize the influence of
loops, we have to look on regions of size comparable to the size of loops. This
motivates the following definitions with $V(x)= \{y\in\bbZ^\nu, |x-y|< R \}$
being an extension of  original neighbourhood $U(x)$. Thus, we enlarge the set
$E(n)$ of
classically excited sites to $\tilde E(n)$, with
$$
\tilde E(n) = \{ x \in \bbZ^\nu : n_{V(x)} \neq g_{V(x)} \text{ for any } g
\in G \}
$$
and we introduce the set $F(n)$ of {\it softly excited sites} by
$$
F(n) = \{ x \in \bbZ^\nu \setminus \tilde E(n) : n_{V(x)} \neq d_{V(x)} \text{
for any } d \in D \} .
$$
Then, for a quantum configuration such that $\bsomega = \check\bsomega$, we
define the new extended boundary
$$
\bsB_{\rm e}(\check\bsomega) = \bigcup_{\tau \in [0,\beta]_\per} \Bigl( \bigl[
\tilde E(\bsn(\tau)) \cup F(\bsn(\tau)) \bigr] \times \tau \Bigr) \bigcup
\bigcup_{i=1}^m \Bigl( \bigl[ \union_{x \in A_i} V(x) \bigr] \times \tau_i
\Bigr) ,
$$
and if $\bsomega \neq \check\bsomega$, we set $\bsB_{\rm e}(\bsomega) =
\bsB_{\rm e}(\check\bsomega)$. Notice that $\bsB(\check\bsomega) \subset
\bsB_{\rm e}(\bsomega)$, since the first set is the union of classical
excitations, quantum transitions and boxes; obviously the classical excitations
and the quantum transitions also belong to $\bsB_{\rm e}(\bsomega)$, and the
boxes being such that their diameter is smaller than $R$ and they contain
$U(x)$-excited sites at each time, they are $V(x)$-excited. Decomposing
$\bsB_{\rm e}(\bsomega)$ into connected components, we get our quantum
contours, namely $\gamma = (B,\bsomega_B)$. Notice that the configuration
$\bsomega_B$ contains actually also  the information determining  which
dominant ground state lies outside $B$. We call the set $B$ the {\it support}
of $\gamma$, $B=\supp \gamma$, and   introduce also its ``truly excited part'',
the {\it core}, $\core\gamma\subset\supp\gamma$, by taking
$\core\gamma=\supp\gamma\bigcap\Bigl(\cup_{\tau \in [0,\beta]_\per} \bigl( 
\tilde E(\bsn(\tau))   \times \tau \bigr) \cup \union_{i=1}^m \Bigl( \bigl[
\union_{x \in A_i} V(x) \bigr] \times \tau_i \bigr)\Bigr)$. Finally, notice
that if the contour is not wrapped around the torus  in its spatial direction,
there exists a space-time configuration $\bsomega^\gamma$ and we have $B =
\bsB_{\rm e}(\bsomega^\gamma)$.

A set of quantum contours $\Gamma = \{ \gamma_1, \dots, \gamma_k \}$ is called
admissible   if there
exists a quantum configuration $\bsomega^{\Gamma}\in\caQ^{\per}_{\Lambda}$ which has
$\Gamma$ as set of quantum contours. Clearly, if it exists, it is unique under
assumption that it contains no loops ($\bsomega^{\Gamma} =
\check\bsomega^{\Gamma}$). 
We use $\caD^{\per}_{\Lambda}$ to denote the set of all collections $\Gamma$
of admissible quantum contours.

Given $\Gamma\in \caD^{\per}_{\Lambda}$, a set of loops $\Xi
=\{\xi_1,\dots,\xi_\ell \}$ is said admissible and compatible with $\Gamma$ if
there exists $\bsomega^{\Gamma\cup\Xi}$ which has $\Xi$ as set of loops and
$\Gamma$ as set of quantum contours (it is also unique whenever it exists). 
More explicitly,
\begin{itemize}
\item two loops $\xi = (B, \bsomega_B, g^\xi_A)$ and $\xi' = (B',
\bsomega_{B'}', g^{\xi'}_{A'})$ are compatible, $\xi \sim \xi'$,  iff $B
\cup B'$ is not connected;
\item using 
$\core\Gamma=\cup_{\gamma\in\Gamma}\core\gamma$, a loop
$\xi = (B,
\bsomega_B, g_A^\xi)$, with 
$B = A
\times [\tau_1,
\tau_2]$, is compatible with $\Gamma$, $\xi \sim
\Gamma$, iff
\ba
& B \cup \core\Gamma \text{ is not connected,} \\
& g^\xi_A = \bsn_A^\Gamma (\tau_1 - 0) =  \bsn_A^\Gamma (\tau_2 + 0);
\end{align}
\item
a collection of loops $\Xi
=\{\xi_1,\dots,\xi_\ell \}$ is  admissible and compatible with $\Gamma$ iff
any two  loops from $\Xi$ are compatible and each loop from $\Xi$ 
is compatible with $\Gamma$.
\end{itemize}
We use $\caD^{\loo}_{\Lambda}(\Gamma)$ to denote the set of all admissible
collections  $\Xi$ that are compatible with $\Gamma$.

The conditions of admissibility and compatibility above can be, for any given set of
transitions  $\{\bsA_1, \dots, \bsA_m \}$,   
formulated as a finite number of restrictions on corresponding transition times 
 $\{\tau_1, \dots, \tau_m \}$. Given the restrictions on admissibility of
$\Gamma\in\caD^{\per}_{\Lambda}$, the restrictions on $\Xi$ to belong to
$\caD^{\loo}_{\Lambda}(\Gamma)$ factorize.
   As a result, the partition function $Z_\Lambda^{\per}$ in \eqref{Zcomp} can be
rewritten in terms of integrations over  $\caD^{\per}_{\Lambda}$ and
$\caD^{\loo}_{\Lambda}(\Gamma)$ [the summation over $\Gamma$ and $\Xi$
accompanied with the integration,  {\it a priori} over the interval
$[0,\beta]$, over times $\tau_i$ of corresponding transitions, subjected to above
formulated restrictions, \cf\eqref{Zexp}].
Furthermore the contribution of $\Gamma \cup \Xi$ factorizes as a contribution
of $\Gamma$ times a product of terms for $\xi \in \Xi$
\cite{BKU,DFF1}\footnote{For spin or boson systems factorization is true simply
because any two operators with disjoint supports commute. In the case of
fermion systems there is an additional sign due to anticommutation relations
between creation and annihilation operators, and factorization is no more
obvious. That it indeed factorizes was nicely proved in Section 4.2 of
\cite{DFF1}.}, we get
\ba
\label{fpart}
Z_\Lambda^{\per} &= \int_{\caD^{\per}_{\Lambda}} \dd\Gamma
\int_{\caD^{\loo}_{\Lambda}(\Gamma)} \dd\Xi \;
\rho^{\per}(\bsomega^{\Gamma\cup\Xi}) \nn\\
&= \int_{\caD^{\per}_{\Lambda}} \dd\Gamma \rho^{\per}(\bsomega^{\Gamma})
\int_{\caD^{\loo}_{\Lambda}(\Gamma)} \dd\Xi
\prod_{\xi \in \Xi} z(\xi) .
\end{align}
Here, using $\{(\bsA_i,\tau_i), i=1, \dots, m\}$ to denote the quantum
transitions of $\Gamma \cup \Xi$, we put
\be
\rho^{\per}(\bsomega^{\Gamma \cup \Xi})= \prod_{i=1}^m \bra{\bsn_{A_i}^{\Gamma \cup
\Xi} (\tau_i-0)} \VAi \ket{\bsn_{A_i}^{\Gamma \cup \Xi}(\tau_i+0)}
\exp\bigl\{-\int_{\bbT_\Lambda}\dd (A,\tau) \Phi_A
(\bsn_A^{\Gamma \cup \Xi}(\tau))\bigr\},
\end{equation}
where $\int_{B}\dd (A,\tau)$ is the shorthand for $\int_0^\beta \dd
\tau\sum_{A: A\times\tau\subset B}$ (used here for $B=\bbT_\Lambda$). 
Similarly for $\rho^{\per}(\bsomega^{\Gamma})$.
Further,
the weight of a loop $\xi = (B^\xi, \bsomega_{B^\xi}, g_A^\xi)$ with the set of
quantum transitions $\{(\bsA_i,\tau_i), i=1, \dots, \ell\}$ and $\bsn^\xi$ the
space-time configuration corresponding to $\bsomega_{B^\xi}$, is
\bm
\label{weightxi}
z(\xi) = \exp \Bigl\{-\int_{B^\xi} \dd (A,\tau)  [\Phi_A(\bsn_A^\xi(\tau)) -
\Phi_A(g_A^\xi)] \Bigr\} \bra{g_{A_1}^\xi} \VAone
\ket{\bsn_{A_1}^\xi(\tau_1+0)}\times\\
\times \bra{\bsn_{A_2}^\xi(\tau_2-0)} \VAtwo \ket{\bsn_{A_2}^\xi(\tau_2+0)}
\dots \bra{\bsn_{A_\ell}^\xi(\tau_\ell-0)} \VAell \ket{g_{A_\ell}^\xi}.
\end{multline}

Given $\Gamma\in \caD^{\per}_{\Lambda}$, the second integral in \eqref{fpart}
is over the collections of the loops that interact only through a condition of
non-intersection. This is the usual framework for applying the cluster
expansion of polymers. The only technical difficulty is that the set of our
loops is uncountable (the loops depend on continuous transition times), and
thus we cannot simply quote the existing literature. Nevertheless, the needed
extension is rather straightforward and often implicitly used.

Given a collection $\bsC = (\xi_1, \dots, \xi_n)$ of loops, we define the
truncated function
\be
\Trfunc(\bsC) = \frac1{n!} \trfunc(\bsC) \prod_{\xi \in \bsC} z(\xi) ,
\end{equation}
with
$$
\trfunc(\bsC) = \trfunc(\xi_1, \dots, \xi_n) = \begin{cases} 1 & \text{if
$n=1$,} \\ \sum_\caG \prod_{e(i,j) \in \caG} \bigl( \indicator{\xi_i \sim \xi_j}
-1 \bigr) & \text{if $n \geq 2$,} \end{cases}
$$
where the sum is over all connected graphs $\caG$ of $n$ vertices. Notice that
$\Trfunc(\bsC) = 0$ whenever $\bsC$ is not a cluster, \ie if the union of the
supports of its loops is not connected. 
We use $\caL_{\Lambda}$ and $\caC_{\Lambda}$  to denote the set of all loops
and clusters, respectively, and use
$\int_{\caC_{\Lambda}}\dd\bsC$ as a shorthand for
$\sum_{n
\geq 1} \int_{\caL_{\Lambda}}\dd\xi_1
\dots \int_{\caL_{\Lambda}}\dd\xi_n$, in obvious meaning.
Whenever  $\Gamma\in \caD^{\per}_{\Lambda}$ is fixed,
we use $\caL_{\Lambda}(\Gamma)$ to denote the set of all loops compatible with
$\Gamma$ and write
$\bsC\in\caC_{\Lambda}(\Gamma)$ whenever the cluster
$\bsC$ contains only loops from $\caL_{\Lambda}(\Gamma)$.
Again,  
$\int_{\caC_{\Lambda}(\Gamma)}\dd\bsC$ is a shorthand for
$\sum_{n\geq 1} \int_{\caL_{\Lambda}(\Gamma)}\dd\xi_1
\dots \int_{\caL_{\Lambda}(\Gamma)}\dd\xi_n$.
Finally, we also need  similar integrals conditioned by the time of the first transition
encountered in the loop $\xi$ or the cluster $\bsC$.  Namely, 
using  $C$ to denote  the support of $\bsC$, \ie the union of the supports of the loops
of $\bsC$,  and $I_C=\{\tau_1(\bsC),\tau_2(\bsC)\}$  to denote  its vertical
projection%
\footnote{Again, if $\tau_1>\tau_2$, the segment
$[\tau_1,\tau_2]\subset[0,\beta]_{\operatorname{per}}$  contains the point
$0\equiv\beta$.},
$I_C=\{\tau\in [0,\beta]_{\operatorname{per}}; \bbZ^\nu\times\tau\cap
C\neq\emptyset\}$, 
we use $\caC_{\Lambda}^{(x,\tau)}$ for the set of all clusters $\bsC\in\caC_{\Lambda}$
with the first transition  time $\tau_1(\bsC)=\tau$, for which their first loop 
$\xi_1 $ with support $B_1=A_1\times[\tau_1(\bsC), \tau_2]$, contains the site $x$,
$A_1\ni x$. Then  $\int_{\caL_{\Lambda}^{(x,\tau)}}\dd\xi$ and
$\int_{\caC_{\Lambda}^{(x,\tau)}}\dd\bsC$ are shorthands for
the corresponding integrals with first transition time fixed
--- formally one replaces $ \int \dd\xi_1$ by 
$ \int \indicator{A_1\ni x}\delta (\tau_1(\xi_1)-\tau)\dd\xi_1$.
With this notation we can formulate the
cluster expansion lemma.

\begin{lemma}[{\bf Cluster expansion}]
\label{lemclexp}

For any $c \in \bbR$, $\alpha_1 < (2R_0)^{-\nu}$, $\alpha_2 < R^{-2\nu}
\Delta_0$ and $\delta > 0$, there exists $\varepsilon_0 > 0$ such
that whenever $\normV \leq \varepsilon_0$ and  $\Gamma\in
\caD^{\per}_{\Lambda}$,  we have the loop cluster expansion, 
\be
\label{klastrrozvoj}
\int_{\caD^{\loo}_{\Lambda}(\Gamma)} \dd\Xi \prod z(\xi) =
\exp\biggl\{\int_{\caC_{\Lambda}(\Gamma)} \dd
\bsC \Trfunc(\bsC)\biggr\}.
\end{equation}
Moreover, the weights of the clusters are  exponentially decaying (uniformly in
$\Lambda$ and $\beta$): 
\be
\label{klastrodhad}
\int_{\caC_{\Lambda}}  \dd
\bsC \indicator{C\ni(x,\tau)} |\Trfunc(\bsC)| \prod_{\xi
\in \bsC} \e{(c - \alpha_1 \log\normV)|A| + \alpha_2 |B|} \leq \delta 
\end{equation}
and
\be
\label{borneclusters}
\int_{\caC_\Lambda^{(x,\tau)}} \dd
\bsC   |\Trfunc(\bsC)| \prod_{\xi
\in \bsC} \e{(c - \alpha_1 \log\normV)|A| + \alpha_2 |B|} \leq \delta 
\end{equation}
for every $(x,\tau)\in\bbT_\Lambda$.
\end{lemma}

{\footnotesize
\begin{proof}

One can follow any standard reference concerning cluster expansions for
continuum systems, for example \cite{Bry}.  We are using here  \cite{Pfi} whose
formulation is closer to our purpose.  Assuming that inequality
\eqref{borneclusters} holds true, we have a finite bound
\be
\sum_{n \geq 1} \frac1{n!} \int_{\caL_\Lambda(\Gamma)^n} \dd\xi_1 \dots
\dd\xi_n |\trfunc(\xi_1, \dots, \xi_n)| \prod_{i=1}^n |z(\xi_i)| \leq \delta
\beta |\Lambda| .
\end{equation}
Lemma \ref{lemclexp} then follows  from Lemma 3.1 of
\cite{Pfi}. Let us turn to the proof of the two inequalities. Let
$$
f(\xi) = |z(\xi)| \e{(c - \alpha_1 \log \normV) |A| + \alpha_2 |B|} .
$$
Skipping the conditions $\xi_j \sim \Gamma $, we define
\be
I_n =  n \Bigl[\int_{\caL_\Lambda} \dd\xi_1 \indicator{B_1 \ni (x,\tau)} + 
\int_{\caL_\Lambda^{(x,\tau)}}\dd\xi_1 \Bigr] \int_{\caL_\Lambda^{n-1}} 
\dd\xi_2 \dots \dd\xi_n |\trfunc(\xi_1, \dots, \xi_n)| \prod_{i=1}^n f(\xi_i)
\end{equation}
(it does not depend on $(x,\tau) \in \bbT_\Lambda$).
The lemma will be completed once we shall have established that  $I_n \leq n!
(\frac12 \delta)^n$ (assuming that $\delta \leq 1$; otherwise, we show that $I_n
\leq n!/2^n$).
From Lemma 3.4 of \cite{Pfi}, we get
\be
|\trfunc(\xi_1, \dots, \xi_n)| \leq \sum_{\caT \text{ tree on $n$ vertices}}
\prod_{e(i,j) \in \caT} \indicator{B_i \cup B_j \text{ connected}} .
\end{equation}
Denoting $d_1, \dots, d_n$ the incidence numbers of vertices $1, \dots, n$, we
first proceed with the integration on the loops $j \neq 1$ for which $d_j = 1$;
in the tree $\mathcal T$, such $j$ shares an edge only with one vertex $i$. 
The incompatibility between $\xi_i$ and $\xi_j$, with $\xi = (B_i,
\bsomega_{B_i}^{(i)}, g_{A_i}^{\xi_i})$, $B_i = A_i \times [\tau_1^{(i)},
\tau_2^{(i)}]$, and similarly for $\xi_j$,  means that either $B_j \cup [A_i
\times \tau_1^{(i)}]$ is connected, or $[A_j \times \tau_1^{(j)}] \cup B_i$ is
connected. Hence, the bound for the integral over the $\xi_j$ that are
incompatible with $\xi_i$ is
\bm
\int_{\caL_\Lambda} \dd\xi_j \indicator{B_j \cap B_i \text{ connected}}
f(\xi_j) \leq 2\nu |A_i| \int_{\caL_\Lambda}\dd\xi_j \indicator{B_j \ni
(x,\tau)} f(\xi_j) + 2\nu |B_i| \int_{\caL_\Lambda^{(x,\tau)}}\dd\xi_j f(\xi_j)
\\
\leq 2\nu \Bigl( |A_i| + \alpha |B_i| \Bigr) \biggl(  \int_{\caL_\Lambda}
\dd\xi_j \indicator{B_j \ni (x,\tau)} f(\xi_j) + \frac1\alpha
\int_{\caL_\Lambda^{(x,\tau)}} \dd\xi_j f(\xi_j) \biggr).
\end{multline}
(The constant $\alpha$ has been introduced in order to match with the
conditions of the next lemma). Then
\bm
I_n \leq n(2\nu)^{n-1} \sum_{\caT \text{ tree of $n$ vertices}} \Bigl[
\int_{\caL_\Lambda}\dd\xi_1 \indicator{B_1 \ni (x,\tau)} +
\int_{\caL_\Lambda^{(x,\tau)}} \dd\xi_1  \Bigr] f(\xi_1) \Bigl( |A_1| + \alpha
|B_1| \Bigr)^{d_1} \\
\prod_{j=2}^n \biggl[ \int_{\caL_\Lambda}\dd\xi_j \indicator{B_j \ni (x,\tau)}
f(\xi_j) \Bigl( |A_j| + \alpha |B_j| \Bigr)^{d_j-1} + \frac1\alpha
\int_{\caL_\Lambda^{(x,\tau)}}\dd\xi_j f(\xi_j) \Bigl( |A_j| + \alpha |B_j|
\Bigr)^{d_j-1} \biggr] .
\end{multline}
Now summing over all trees, knowing that the number of trees with $n$
vertices and incidence numbers $d_1, \dots, d_n$ is equal to
$$
\frac{(n-2)!}{(d_1-1)! \dots (d_n-1)!} \leq \frac{(n-1)!}{d_1! (d_2-1)! \dots
(d_n-1)!} ,
$$
we find a bound
\be
I_n \leq n! (2\nu)^{n-1} (1+\alpha) \biggl[ \int_{\caL_\Lambda}\dd\xi
\indicator{B \ni (x,\tau)} f(\xi) \e{|A| + \alpha |B|} + \frac1\alpha
\int_{\caL_\Lambda^{(x,\tau)}} \dd\xi  f(\xi) \e{|A| + \alpha |B|} \biggr]^n .
\end{equation}
We conclude by using the following lemma which implies that the quantity
between the brackets is small.

\end{proof}
}

\begin{lemma}
\label{lemboundxi}
Let $\alpha_1 < (2R_0)^{-\nu}$ and $\alpha_2 < R^{-2\nu} \Delta_0$. For any $c \in
\bbR$ and $\delta > 0$, there exists $\varepsilon_0 > 0$ such that whenever $\normV \leq
\varepsilon_0$ the following inequality holds true,
$$
\int_{\caL_\Lambda} \dd\xi \indicator{B \ni (x,\tau)} |z(\xi)| \e{(c - \alpha_1
\log\normV) |A| + \alpha_2 |B|} + \int_{\caL_\Lambda^{(x,\tau)}} \dd\xi 
|z(\xi)| \e{(c - \alpha_1 \log\normV) |A| + \alpha_2 |B|} \leq \delta ,
$$
where $(x,\tau)$ is any space-time site of $\bbT_\Lambda$.
\end{lemma}

{\footnotesize
\begin{proof}

Let us first consider the integral over $\xi$ such that its box contains a
given space-time site. We denote by $\ell_1$ the number of quantum transitions
of $\xi$ at times bigger than $\tau$, and $\ell_2$ the number of the other
quantum transitions. The integral over $\xi$ can be done by summing over
$(\ell_1 + \ell_2)$ quantum transitions $\bsA_1^1, \dots, \bsA^1_{\ell_1},
\bsA_1^2, \dots, \bsA^2_{\ell_2}$, by summing over $(\ell_1 + \ell_2)$
configurations $n^{i,j}_{A^i_j}$, and by integrating over times $\tau_1^1 < \dots <
\tau^1_{\ell_1}$, $\tau_1^2 < \dots < \tau^2_{\ell_2}$. Let us do the change of
variables $\tilde\tau_1^1 = \tau_1^1 - \tau$, $\tilde\tau_2^1 = \tau_2^1 -
\tau^1_1$, \dots, $\tilde\tau^1_{\ell_1} = \tau^1_{\ell_1} -
\tau^1_{\ell_1-1}$, and $\tilde\tau_1^2 = \tau - \tau_1^2$, \dots,
$\tilde\tau^2_{\ell_2} = \tau^2_{\ell_2-1} - \tau^2_{\ell_2}$. Then we can
write the following upper
bound
\bm
\label{boundxi}
\int_{\caL_\Lambda} \dd\xi \indicator{B \ni (x,\tau)} |z(\xi)| \e{(c - \alpha_1
\log\normV) |A| + \alpha_2 |B|} \leq \sum_{\ell_1, \ell_2 \geq 1} \sumthree{\bsA^1_1,
\dots
\bsA^2_{\ell_2}}{\cup_{i,j} \bar A_j^i = A \ni x}{A \text{ connected}}
\sum_{n^{1,1}_{A^1_1}, \dots, n^{2,\ell_2}_{A^2_{\ell_2}} \notin G_A}
\int_0^\infty \dd\tilde\tau^1_1 \dots \dd\tilde\tau^2_{\ell_2} \\
\prod_{i =1,2} \prod_{j=1}^{\ell_i} |\bra{n_A^{i,j}} \bsV_{\!\!\! \bsA^i_j}
\ket{n_A^{i,j+1}}| \e{(c - \alpha_1 \log\normV) |\bar A^i_j|}
\e{-\tilde\tau^i_j \sum_{A' \subset A} [\Phi_{A'}(n^{i,j}_{A'}) -
\Phi_{A'}(g_{A'})]} \e{\tilde\tau_j^i R^\nu \alpha_2}
\end{multline}
where $g_A \in G_A$ is the configuration in which the loop $\xi$ is immersed
(if the construction does not lead to a possible loop, we find a bound by
picking any $g_A \in G_A$). Remark that we neglected a constraint on the sum
over configurations, namely $n_A^{1,1} = n_A^{2,1}$. It is useful to note that
the sums over $\ell_1,\ell_2$ and over the quantum transitions are finite,
otherwise they cannot constitute a loop.

Using the definition \eqref{defnorm} of the norm of a quantum interaction, we
have
$$
|\bra{n_A'} \VA \ket{n_A}| \leq \normV^{|A|} .
$$
Furthermore
$$
\sum_{A' \subset A} [\Phi_{A'}(n^{i,j}_{A'}) - \Phi_{A'}(g_{A'})] \geq R^{-\nu}
\Delta_0
$$
as claimed in Property \eqref{boundphi}. Hence we have, since the number of
configurations on $A$ is bounded with $S^{|A|}$,
\be
\label{boundxi2}
\int_{\caL_\Lambda} \dd\xi \indicator{B \ni (x,\tau)} |z(\xi)| \e{(c - \alpha_1
\log\normV) |A| + \alpha_2 |B|}
\leq \sum_{\ell_1, \ell_2 \geq 1} \sumthree{\bsA^1_1, \dots
\bsA^2_{\ell_2}}{\cup_{i,j} \bar A_j^i = A \ni x}{A \text{ connected}} \prod_{i
=1,2} \prod_{j=1}^{\ell_i} \frac{\bigl[ \normV^{1 - \alpha_1 (2R_0)^\nu} S \e{c(2R_0)^\nu}
\bigr]^{|A^i_j|}}{R^{-\nu} \Delta_0 - R^\nu \alpha_2} .
\end{equation}
This is a small quantity since the sums are finite, by taking $\normV$ small
enough.
Now we turn to the second term, namely
$$
\int_{\caL_\Lambda^{(x,\tau)}} \dd\xi |z(\xi)|
\e{(c - \alpha_1
\log\normV) |A| + \alpha_2 |B|}.
$$
The proof is similar; we first sum over the number of transitions $\ell$, then
over $\ell$ transitions $\bsA_1, \dots \bsA_\ell$ with $A = \cup_i \bar A_i \ni
x$, $A$ connected. Then we choose $\ell-1$ intermediate configurations. Finally,
we integrate over $\ell-1$ time intervals. The resulting equation looks very
close to \eqref{boundxi} and is small for the same reasons.

\end{proof}
}	

Now, we single out the class of {\it small clusters}. Namely,  a cluster is
small if the sequence of its quantum transitions belongs to the list $\caS$. To
be more precise, we have to specify  the order of transitions: considering a
cluster $ \bsC\equiv(\xi_1, \dots , \xi_k)$ and using $S(\xi^{(\ell)})$,
$\ell=1, \dots, k$,  to denote the sequence of quantum transitions of the loop
$\xi^{(\ell)}=(B^{(\ell)}, \bsomega_{B^{(\ell)}}, g_A^{\xi^{(\ell)}})$, 
$S(\xi^{(\ell)})\equiv S(B^{(\ell)}, \bsomega_{B^{(\ell)}})$, we take the
sequence $S(\bsC)$ obtained by combining the sequences $S(\xi^{(1)}), \dots ,
S(\xi^{(k)})$ in this order. A cluster $\bsC$ is said to be {\it small} if
$S(\bsC)\in \caS$, it is {\it large} otherwise. We use $\caC_{\Lambda}^{\sma}$
to denote the set of all small clusters on the torus $\bbT_{\Lambda}$.

The local contribution to the energy at time  $\tau$,  when the system is in a
state $\bsn_{A}(\tau)$, is $\Phi_A(\bsn_{A}(\tau))$. Similarly, we will
introduce the local contribution of loops (and small clusters of loops)  in the
expansion of the partition function --- the effective potential
$\Psi_A^\beta(\bsn_{A}(\tau))$. The latter is a local quantity in the sense
that it depends on $\bsn$ only on the set $A$ at time
$\tau$. An explicit  expression of $\Psi_A^\beta(g_{A})$ with $g \in
G$ is, in terms  of small clusters,
\be
\label{effpottemp}
\Psi_A^\beta(g_{A}) \isdefby -\int_{\caC_{\Lambda}^{\sma}}\dd \bsC
\frac{\Trfunc(\bsC)}{|I_C|}
\indicator{\bsC \sim g_A, A_C = A, I_C \ni 0} .
\end{equation}
Here, again, $C$ is the support of $\bsC$,  $A_C$  its horizontal projection 
onto $\bbZ^\nu$, $A_C=\{x\in\bbZ^\nu; x\times[0,\beta]_{\operatorname{per}}\cap
C\neq\emptyset\}$, and  $I_C$ its vertical  projection, $|A_C|$ and $|I_C|$
their corresponding areas, and the condition $\bsC \sim g_{A}$ means that each
loop of $\bsC$ is immersed in the ground state $g$.  Notice that ``horizontal
extension'' of any small cluster is at most $R$: if $\bsC$ is a small cluster,
$\diam(A_C) \leq R$. The definitions of Section \ref{secdefpot} are now clear, 
once we identify the effective potential $\Psi$ defined in \eqref{defeffpot} as
the limit $\beta \to \infty$ of \eqref{effpottemp}. Namely,
$$
\Psi =\lim_{\beta \to \infty} \Psi^\beta .
$$

Our assumptions in Section \ref{secstab} concern the limit $\beta \to \infty$ of
the effective potential, but at non zero temperature we have to work with
$\Psi^\beta$. To trace down the difference, we introduce
$\psi^\beta = \Psi^\beta - \Psi$. Notice that
\eqref{effpottemp} implies $\Psi_A^\beta(n_A) = 0$ whenever $n_A \notin G_A$
or $\diam A < 2R_0$.

Recalling that if
$C \subset \bbT_\Lambda$, $\tilde C$ is the smallest box containing $C$, we
 introduce, for
any cluster $\bsC \in \caC_{\Lambda}^{\sma}$,   the function
\be
\label{PhiTCGamma}
\Trfunc(\bsC; \Gamma) =  \frac{\Trfunc(\bsC)}{|I_C|}\int_0^\beta \dd\tau
 \indicator{  I_C\ni \tau}\Bigl( \indicator{\bsC \sim \Gamma} -
\indicator{\bsn_{A_C}^\Gamma(\tau) \in G_{A_C},\bsC\sim\bsn_{A_C}^\Gamma(\tau)  }
\Bigr) .
\end{equation}
Here, the first indicator function in the parenthesis singles out the clusters
whose each loop is compatible with $\Gamma$, while the second indicator
concerns the clusters for which $\bsn_{A_C}^\Gamma(\tau)\in G_{A_C}$ and  each
their loop is immersed in the configuration  $\bsn_A^\Gamma(\tau)$ (extended as
a constant to all the time interval $I_C$). Observing that
$\Trfunc(\bsC;\Gamma)=0$ whenever  $\tilde C\cap\core\Gamma=\emptyset$, we
split the integral over small clusters into its bulk part expressed in terms of
the effective potential  and boundary terms ``decorating'' the quantum contours
from $\Gamma$.

\begin{lemma}
\label{lemdecclusters}
For any fixed  $\Gamma\in \caD_{\Lambda}$, one has
\bm
\int_{\caC_{\Lambda}^{\sma}(\Gamma)} \dd\bsC \Trfunc(\bsC) =
-\int_{\bbT_\Lambda} \dd(A,\tau) \Psi_A(\bsn_A^\Gamma(\tau)) -
\int_{\bbT_\Lambda} \dd(A,\tau) \psi_A^\beta(\bsn_A^\Gamma(\tau)) 
+ \int_{\caC_{\Lambda}^{\sma}}
\dd\bsC \Trfunc(\bsC; \Gamma) .\nn
\end{multline}
The term $\Trfunc(\bsC;\Gamma)$ vanishes whenever 
$\tilde C\cap\core\Gamma=\emptyset$.
\end{lemma}

{\footnotesize
\begin{proof}

To get the equality of integrals, it is enough  to rewrite
\be
\int_{\caC_{\Lambda}^{\sma}(\Gamma)} \dd\bsC \Trfunc(\bsC) = 
\int_{\caC_{\Lambda}^{\sma}} \dd\bsC \Trfunc(\bsC) 
\indicator{ \bsC \sim \Gamma}
\end{equation}
and
\be
-\int_{\bbT_\Lambda} \dd(A,\tau) \Psi_A^\beta(\bsn_A^\Gamma(\tau))=
\int_{\caC_{\Lambda}^{\sma}} \dd\bsC \Trfunc(\bsC) \int_0^\beta \dd\tau
\indicator{\bsn_{A_C}^\Gamma(\tau) \in G_{A_C},\bsC\sim\bsn_{A_C}^\Gamma(\tau)}.
\end{equation}
Moreover, whenever $\tilde C\cap\core\Gamma=\emptyset$, the configuration
 $\bsn_{A_C}^\Gamma(\tau)$ belongs to $G_{A_C}$, and  it is constant, for all $\tau\in
I_C$. Under these circumstances,   the condition 
$\bsC \sim \Gamma$ is equivalent to 
$\bsC\sim\bsn_{A_C}^\Gamma(\tau)$ and the right hand side of 
\eqref{PhiTCGamma} vanishes.

\end{proof}
} 

Whenever $\Gamma\in \caD_{\Lambda}$ is fixed, let $W_d(\Gamma) \subset
\bbT_\Lambda$ be the set of space-time sites in the state
$d$, \ie
$$
W_d(\Gamma) = \{ (x,\tau) \in \bbT_\Lambda : \bsn_{V(x)}^\Gamma(\tau) =
d_{V(x)}  \}.
$$
Notice that 
$$
\bbT_\Lambda = \supp\Gamma \union \union_{d \in D} W_d(\Gamma) ;
\hspace{4mm} W_d(\Gamma) \cap W_{d'}(\Gamma) = \emptyset \text{ if } d \neq d' ,
$$
and the set $\supp\Gamma \cap W_d(\Gamma)$ is of measure zero
(with respect to the measure $\dd(x,\tau)$ on $\bbT_\Lambda$).
Let us recall that the equivalent potential $\Upsilon$ satisfies the equality
$\sum_{A\subset\Lambda}\Upsilon_A(n_A)=\sum_{A\subset\Lambda}
(\Phi_A(n_A)+\Psi_A(n_A))$ for any configuration $n$ on the torus $\Lambda$ and that
we defined $e(d)=\sum_{A\ni 0}\frac{\Upsilon_A(d_A)}{|A|}$ for every $d\in D$.

\begin{lemma}
\label{lemfpart}
The partition function \eqref{fpart} can be rewritten as
$$
Z^{\per}_\Lambda = \int_{\caD^{\per}_\Lambda}\dd\Gamma \prod_{d \in D}
\e{-|W_d(\Gamma)| e(d)}
\prod_{\gamma \in \Gamma} z(\gamma) \e{\caR(\Gamma)} .
$$
Here the weight $z(\gamma)$ of a quantum contour $\gamma = (B,\bsomega_B)$ with
the sequence of transitions $(\bsA_1, \dots, \bsA_m)$ at times $(\tau_1, \dots,
\tau_m)$ is
\be
\label{defz}
z(\gamma) = \prod_{i=1}^m \bra{\bsn_{A_i}^\gamma(\tau_i-0)} \VAi
\ket{\bsn_{A_i}^\gamma(\tau_i+0)} \exp \Bigl\{ -\int_B \dd(x,\tau) 
e_x^{\Upsilon}(\bsn^\gamma(\tau)) 
\Bigr\}.
\end{equation}
The rest $\caR(\Gamma)$ is given by
\be
\label{deffrR}
\caR(\Gamma) =
\int_{\caC_{\Lambda}(\Gamma)\setminus\caC_{\Lambda}^{\sma}(\Gamma)}
\dd\bsC
\Trfunc(\bsC) -
\int_{\bbT_\Lambda} \dd(A,\tau) \psi^\beta_A(\bsn_A^\Gamma(\tau)) +
\int_{\caC_{\Lambda}^{\sma}} \dd\bsC
\Trfunc(\bsC; \Gamma) .
\end{equation}
\end{lemma}

{\footnotesize
\begin{proof}

Using the Lemmas \ref{lemclexp} and \ref{lemdecclusters} to substitute in \eqref{fpart}
the contribution of  loops by the action of the effective potential, we get
\be
Z^{\per}_\Lambda = \int_{\caD^{\per}_\Lambda}\dd\Gamma \Bigl\{ \prod_{i=1}^m
\bra{n_{A_i}^\Gamma(\tau_i-0)} \VAi \ket{n_{A_i}^\Gamma(\tau_i+0)}
\Bigr\} \exp\Bigl\{ -\int_{\bbT_\Lambda} \dd(A,\tau) (\Phi_A
(\bsn_A^\Gamma(\tau))  + \Psi_A(\bsn_A^\Gamma(\tau)) )\Bigr\} \e{\caR(\Gamma)} .
\end{equation}
Replacing $\Phi + \Psi$ by the physically equivalent potential $\Upsilon$, we get 
\be
\label{replacement}
Z^{\per}_\Lambda = \int_{\caD^{\per}_\Lambda}\dd\Gamma \Bigl\{ \prod_{i=1}^m
\bra{n_{A_i}^\Gamma(\tau_i-0)} \VAi \ket{n_{A_i}^\Gamma(\tau_i+0)}
\Bigr\} 
\exp \biggl\{ -\int_{\supp\Gamma} \dd(x,\tau) 
e_x^{\Upsilon}(\bsn^\gamma(\tau)) 
\biggr\} \prod_{d \in D} \e{-e(d)|W_d(\Gamma)|} \e{\caR(\Gamma)} .
\end{equation}
We get our lemma by observing that the product over quantum transitions and the
first exponential factorize with respect to the quantum contours, as it was the
case for the loops (for fermions the sign is arising because of anticommutation
relations also factorize; we again refer to \cite{DFF1} for the proof).

\end{proof}
} 

Our goal is to obtain a classical lattice system in $\nu+1$ dimensions. Thus we
introduce a discretization of the continuous time direction, by choosing
suitable parameters $\tilde\beta > 0$ and $N \in \bbN$ with $\beta = N
\frac{\tilde\beta}{\Delta}$.\footnote{Remark the difference from \cite{BKU};
here the vertical length of a unit cell $\tilde\beta / \Delta$ depends on
$\normV$, since so does the quantum Peierls constant  $\Delta$.} Setting
$\bbL_\Lambda$ to be the  $(\nu +1)$-dimensional discrete torus $\bbL_\Lambda =
\Lambda \times \{ 0, 1, \dots, N-1 \}^\per$ --- let us recall that $\Lambda$
has periodic boundary conditions in all spatial directions --- and  using
$C(x,t) \subset \bbR^{\nu+1}$ to denote, for any $(x,t)\in \bbL_\Lambda$,  the
cell centered in $(x, \frac{\tilde\beta}\Delta t)$ with vertical length
$\tilde\beta/\Delta$, we have $\bbT_\Lambda = \union_{(x,t)\in \mathbb
L_\Lambda} C(x,t)$. 

For any $M\subset \bbL_\Lambda$, we set $C(M)$ to be the union of all cells
centered at sites of $M$, $C(M) =
\cup_{(x,t)
\in M} C(x,t) \subset \bbT_\Lambda$.  
Conversely, if $B \subset \bbT_\Lambda$, we take $M(B) \subset \bbL_\Lambda$ to be
the smallest set such that $C(M(B)) \supset B$.
Given a connected%
\footnote{Connectedness in $\bbL_\Lambda$ is meant in standard way via nearest
neighbours.} set $M\subset
\bbL_\Lambda$ and a collection of quantum contours 
$\Gamma\in\caD_\Lambda^{\per}$, we define
\bm
\label{defvarphi}
\varphi(M;\Gamma) =
\int_{\caC_{\Lambda}(\Gamma)\setminus\caC_{\Lambda}^{\sma}(\Gamma)}
\dd\bsC\indicator{M(C) =M}
\Trfunc(\bsC) +\\ + \int_{\caC_{\Lambda}^{\sma}}
\dd\bsC \indicator{M(C) =M,C\not\subset C(\supp \Gamma)}\Trfunc(\bsC; \Gamma) -
\int_{M(A \times \tau) = M} \dd(A,\tau)
\psi_A^\beta(\bsn_A^\Gamma(\tau))
\end{multline}
and
\be
\label{defRA}
\tilde\caR(\Gamma) = 
\int_{\caC_{\Lambda}^{\sma}} 
\dd\bsC\indicator{ C\subset C(\supp \Gamma)}
\Trfunc(\bsC; \Gamma) .
\end{equation}
We have separated the contributions of the small clusters inside
$C(\supp\Gamma)$, because they are not necessarily a small quantity, and it is
impossible to expand them. On the contrary, $\varphi(M;\Gamma)$ is small, and
hence it is natural to write
\be
\label{exprest}
\e{\caR(\Gamma)} = \e{\tilde\caR(\Gamma)} \sum_{\caM} \prod_{M \in \caM} \Bigl(
\e{\varphi(M;\Gamma)} - 1 \Bigr) ,
\end{equation}
with  the sum running over all collections $\caM$ of connected subsets of
$\bbL_\Lambda$.

Let $\supp\caM = \cup_{M \in \caM} M$.  Given a set of
quantum contours $\Gamma\in\caD_\Lambda^{\per}$ and a collection
$\caM$, we introduce   contours on $\bbL_\Lambda$ by
decomposing the set 
$M(\supp\Gamma)\cup \supp\caM$  into connected components [notice that
if $(x,t) \notin M(\supp\Gamma)\cup \supp\caM$, then $C(x,t) \subset \cup_{d \in D}
W_d(\Gamma)$].
Namely, a {\it contour} $Y$ is a pair $(\supp Y, \alpha_Y)$ where $\supp
Y \subset \bbL_\Lambda$ is a (non-empty) connected subset of $\bbL_\Lambda$,
and $\alpha_Y$ is a labeling of connected components $F$ of $\partial C(\supp
Y)$, $\alpha_Y(F) = 1, \dots, r$.  We write $|Y|$
for the length (area) of the contour $Y$, \ie the number of sites in $\supp Y$. A set
of contours $\caY = \{ Y_1, \dots, Y_k \}$ is {\it admissible} if the contours
are mutually disjoint and if the labeling is constant on the boundary of each
connected component of $\bbT_\Lambda \setminus \cup_{Y \in \caY} C(\supp
Y)$.  Finally, given an admissible
set of contours $\caY$, we define $\caW_d(\caY)$ to be the union of all
connected components $M$ of $\bbL_\Lambda \setminus \cup_{Y \in \caY} \supp Y$
such that $C(M)$ has label $d$ on its boundary.

Consider now  any configuration $\bsomega\in\caQ_\Lambda^{\per}$ 
yielding, together with a collection $\caM$, a fixed set of contours $\caY$.
Summing over all such configurations $\bsomega$ and collections $\caM$,
we get the weight to be attributed to the set  $\caY$. 
Let $\Gamma^{\bsomega}$ be the collection of quantum contours corresponding to 
$\bsomega$,
$\cup_{Y\in\caY} \supp Y =M(\supp\Gamma^{\bsomega})\cup \supp\caM$.
Given that the configurations 
$\bsomega$ are necessarily constant with no transition on 
$\bbT_\Lambda\setminus C( M(\supp\Gamma^{\bsomega})\cup \supp\caM)$, we easily
see that the weight factor splits into product of weight factors of single contours
$Y\in\caY$.  Namely,  for the  weight $\frz$ of a contour $Y$
we get  the expression
\bm
\label{deffrz}
\frz(Y) = \int_{\caD_\Lambda^{\per}(Y)}\dd\Gamma 
\prod_{\gamma \in \Gamma} z(\gamma) \prod_{d \in D} \e{-e(d) |W_d(\Gamma) \cap
C(\supp Y)|} \e{\tilde\caR(\Gamma)} \\
\sum_\caM \indicator{M(\supp \Gamma) \union \supp\caM =
\supp Y} \prod_{M \in \caM} \Bigl( \e{\varphi(M; \Gamma)} - 1 \Bigr) ,
\end{multline}
where $\caD_\Lambda^{\per}(Y)$ is the set of collections $\Gamma$ of quantum contours
compatible with $Y$, $\Gamma\in \caD_\Lambda^{\per}(Y)$ if $\supp\Gamma\subset
\supp Y$ and the labels on  the boundary of $\supp \Gamma$ match with labels of $Y$.
Thus, we can finally rewrite the partition function  in a form that
agrees with the standard  Pirogov-Sinai setting,  namely
\be
\label{periodparticni}
Z_\Lambda^{\per} = \sum_\caY \prod_{d \in D} \e{-\frac{\tilde\beta}\Delta e(d)
|\caW_d(\caY)|} \prod_{Y \in \caY} \frz(Y) ,
\end{equation}
with the sum being over all admissible sets of contours on $\bbL_\Lambda$.

In the next section we will evaluate the decay rate of  contours  weights in a
preparation to apply, in Section \ref{secexpobs}, the Pirogov-Sinai theory to prove
Theorems \ref{thmlimit}, \ref{thmphases}, and  \ref{thmphasediagram}.

\section{Exponential decay of the weight of the contours}
\label{secbounds}

In this section we show that the weight $\frz$ has exponential decay with
respect to the length of the contours. We begin by a lemma proving that the
contribution of $\caM$ is small, that we shall use in Lemma \ref{lemboundweight}
below for the bound of $\frz$.

\begin{lemma}
\label{lemboundrest}
Under the Assumptions \ref{assdiagpot}--\ref{assnoquinst}, for any $c <
\infty$  there exist constants $\beta_0$, $ \tilde\beta_0 < \infty$, and
$\varepsilon_0 > 0$ such that for any $\beta \geq \beta_0$, $\tilde\beta_0 \leq
\tilde\beta < 2\tilde\beta_0$, and $\normV \leq \varepsilon_0$, one has
$$
\sum_{M \ni (x,t)} \bigl| \e{\varphi(M; \Gamma)} - 1 \bigr| \e{c|M|} \leq 1 
$$
for any contour $Y$ and any set of quantum
contours $\Gamma\in \caD_\Lambda^{\per}(Y)$.  
  \end{lemma}

{\footnotesize
\begin{proof}
We show that
$$
\sum_{M \ni (x,t)} \bigl| \varphi(M; \Gamma) \bigr| \e{c|M|} \leq 1 .
$$
This implies that $|\varphi(M; \Gamma)| \leq 1$ and consequently Lemma
\ref{lemboundrest} holds --- with a slightly smaller constant $c$.

Let us consider separately, in \eqref{defvarphi},  the three terms on the right hand
side:
{\it (a)} the integral over big clusters, {\it (b)} the integral over small clusters, and
{\it (c)} the expression involving $\psi^\beta$.

{\it (a) Big clusters.} 
Our aim is to estimate
$$
J = \sum_{M\ni (x,t)} \e{c|M|}  \int_{\caC_\Lambda(\Gamma)\setminus
\caC_\Lambda^{\sma}(\Gamma) } \dd\bsC \indicator{M(C) =
M}\bigl|\Trfunc(\bsC)\bigr| .
$$

Since $M(C) = M$ and $M \ni (x,t)$, the cell $C(x,t)$  either intersects a
quantum transition of $\bsC$, or  it  is contained in a box $B$ belonging to a loop
of $\bsC$ (both possibilities may occur at the same time). In the first case we
start the integral over clusters by choosing the  time for the first quantum
transition, which yields a factor $\tilde\beta / \Delta$. In the second case we
simply integrate over all loops containing the given site. 
In the same time,
given a cluster $\bsC =
(\xi_1, \dots, \xi_n)$, $\xi_i = (B_i, \bsomega_{B_i}^{(i)}, g_{A_i}^{\xi_i})$
and $B_i = A_i \times [\tau_1^{(i)}, \tau_2^{(i)}]$, 
the condition  $M(C) = M$ implies that
\be
\label{Cbig}
\sum_{i=1}^n \Bigl\{ |A_i| + \frac{\Delta}{\tilde\beta} |B_i| \Bigr\} \geq |M|
.
\end{equation}
Using it to bound $|M|$, we get the estimate
\be
\label{Jodhad}
J \leq  \frac{\tilde\beta}\Delta \int_{\caC_\Lambda^{(x,\tau)}\setminus
\caC_\Lambda^{\sma}} \dd\bsC  |\Trfunc(\bsC)| \prod_{\xi \in \bsC} \e{c|A| +
c\frac{\Delta}{\tilde\beta} |B|} + \int_{\caC_\Lambda\setminus
\caC_\Lambda^{\sma}} \dd\bsC \indicator{C\ni(x,\tau)} |\Trfunc(\bsC)|
\prod_{\xi \in \bsC} \e{c|A| + c\frac{\Delta}{\tilde\beta} |B|} .
\end{equation}

Taking, in Lemma 4.1, the constant $c$ as above as well as $\alpha_1= \frac12
(2R_0)^{-\nu}$, $\alpha_2= c \Delta / \tilde\beta$,  $\delta=1$, and choosing
the corresponding $\varepsilon_0(c,\alpha_1,\alpha_2,\delta)$, we can bound 
the second term  of \eqref{Jodhad}, for any $\normV\leq \varepsilon_0$,  with
the help of  \eqref{klastrodhad} once $\tilde\beta$ is chosen large enough to
satisfy
\be
\label{vlnkaodhad}
\frac{\tilde\beta}{\Delta}> \frac{c}{\Delta_0}R^{2\nu}.
\end{equation}

To estimate the first term of \eqref{Jodhad}, we first consider  the
contribution of those clusters  for which 
$$
\frac{\tilde\beta}{\Delta} \leq  \prod_{\xi \in \bsC} \normV^{\frac12
(2R_0)^{-\nu} |A|} .
$$
Applying it together with \eqref{vlnkaodhad} we can directly use the bound
\eqref{borneclusters}.

Thus it remains to estimate the contribution of those terms for which
\be
\label{cond}
\frac2{(2R_0)^\nu} \sum_{\xi \in \bsC} |A| < \frac{\log(\Delta/\tilde\beta)}
{\log\normV} .
\end{equation}
Let us first fix $\tilde\beta$ and $\varepsilon_0\leq
\varepsilon_0(c,\alpha_1,\alpha_2,\delta)$  with the constants $c$, $\alpha_1$,
$\alpha_2$, and $\delta$ as above, so that 
\be
\frac{\tilde\beta}{\varepsilon_0}> \frac{c}{\Delta_0}R^{2\nu}
\end{equation}
and, in the same time,
\be
\tilde\beta\leq \varepsilon_0^{k-2{k'}{(2R_0)^{-\nu}}}
\end{equation}
for a suitable large $k'$ (we also assume that $\varepsilon_0 \leq 1$).  Here
$k$ is the constant that appears in Assumption \ref{assPeiprop},
$\Delta(\normV)\geq \normV^k$.  Observing further that $\Delta(\normV)$ can be
taken to increase with $\normV$ (one can always consider a weaker lower bound
$\Delta$ when taking smaller $\normV$), we conclude that  \eqref{vlnkaodhad},
as well as the condition
$$
\tfrac12 (2R_0)^\nu\frac{\log(\Delta/\tilde\beta)}{\log\normV} \leq k',
$$
are satisfied for every $\normV\leq\varepsilon_0$. Thus, it suffices to find an upper
bound to
\be
J' = \frac{\tilde\beta}\Delta \int_{\caC_\Lambda^{(x,\tau)} \setminus
\caC_\Lambda^\sma} \dd\bsC |\Trfunc(\bsC)| \indicator{\sum_{\xi \in \bsC} |A| <
k'} .
\end{equation}
The main problem in estimating this term stems from  the factor
$1/\Delta$ that may be large if
$\normV$ is small. Thus, to have a bound valid for all small $\normV$,
some terms, coming from the integral, that would suppress  this factor  must be
displayed.

The condition $\sum_{\xi \in \bsC} |A| < k'$ will be used several times by
applying  its obvious  consequences: (i) the number of loops in $\bsC$ is
smaller than $k'$, (ii) the number of transitions for each loop is smaller than
$k'$, (iii) each transition $\bsA$ is such that $|A| < k'$, and (iv) the
distance between each transition and $x$ is smaller than $k'$.

Furthermore, we use Assumption \ref{asscompl} to bound the contribution of the
transitions of $\bsC$; recalling the definition \eqref{weightxi} of the weight
of $\xi$, we have, for any large $\bsC$,
\ba
\prod_{\xi \in \bsC} |z(\xi)| &\leq  b_1(\normV) \Delta\prod_{\xi \in \bsC} \exp
\Bigl\{ -\int_B \dd(A,\tau) [\Phi_A(n_A^\xi(\tau)) - \Phi_A(g_A^\xi(\tau))]
\Bigr\} \nn\\
&\leq  b_1(\normV)\Delta \prod_{\xi \in \bsC} \e{-R^{-2\nu} \Delta_0 |B|} .
\end{align}
In the last inequality we used  Assumption \ref{assgap} in the form of the bound
\eqref{boundphi} as well as the lower bound
$|\tau_2-\tau_1| = \frac{|B|}{|A|}\geq \frac{|B|}{R^\nu}$
 for the support $B=A\times [\tau_1,\tau_2]$ of the loop $\xi$.

For any $\xi\in\bsC=(\xi_1,\dots\xi_n)$, let  $\tau$ be  the time at which the
first transition in $C$ occurs (we assume that it happens for the ``first''
loop $\xi_1$) and  $\tau^\xi$ be such that $\tau + \tau^\xi$ is the time at
which the first transition in $\xi$ occurs ($\tau^{\xi_1} = 0$).  Referring to
the condition (i) on the number of loops in   $\bsC$, we get the inequality
$$
\sum_{\xi \neq \xi_1} |\tau^\xi| \leq k' \sum_\xi |B|,
$$
and thus also
$$
1 \leq \prod_\xi\e{-\frac{\Delta_0}{2k' R^{2\nu}}  |\tau^\xi|}\prod_\xi\e{  \frac12
R^{-2\nu} \Delta_0  |B|} .
$$
Integrating now over the time of the first transition for each $\xi \in \bsC$, $\xi \neq
\xi_1$, and taking into account that
$|\trfunc(\xi_1,
\dots,
\xi_n)| \leq n^{n-2}$,  we get 
\be
J' \leq \tilde\beta b_1(\normV) \sum_{n=1}^{k'} \frac{n^{n-2}}{(n-1)!} \Bigl(
\frac{2k' R^{2\nu}}{\Delta_0} \Bigr)^{n-1}
\Bigl\{\int_{\caL_\Lambda^{(x,\tau)}} \dd\xi \e{-\frac12 R^{-2\nu} \Delta_0
|B|} \indicator{\xi : k'}\Bigr\}^{n}.
\end{equation}
Here the constraint $\indicator{\xi_i : k'}$ means that the loop $\xi_i$
satisfies the conditions (ii)--(iv) above. We have then a finite number of
finite terms, the contribution of which is bounded by a fixed number $K < \infty$
(depending on $\varepsilon_0$, $\tilde\beta$, and $k'$).
Thus $J' \leq \tilde\beta b_1(\normV)K$ which can be made small by taking $\normV$
sufficiently small.

{\it (b) Small clusters.} Let us first notice that $|\Trfunc(\bsC; \Gamma)|
\leq |\Trfunc(\bsC)|$, and since $M(C) = M$,  inequality \eqref{Cbig} is valid.
Moreover $C$ must contain at least one of the two boundary points  $(y,
t\frac{\tilde\beta}{\Delta}\pm\frac{\tilde\beta}{2\Delta})$ of some cell 
$C(y,t)$ for which $\dist(x,y)\leq R$. Indeed,   given that $\bsC$ is small and
in the same time  $\tilde C\cap\core \Gamma\neq \emptyset$ (\cf Lemma
\ref{lemdecclusters}),  this is the only way to satisfy also $C\not\subset
C(\supp \Gamma)$ [\cf  \eqref{defvarphi}]. Thus it suffices to use again  
\eqref{klastrodhad}   and \eqref{vlnkaodhad} to estimate
$$
(2R)^\nu \int_{\caC_\Lambda^{\sma}} \dd\bsC \indicator{C \ni (x,\tau)}
|\Trfunc(\bsC)| \prod_{\xi \in \bsC} \e{c|A| + c \frac\Delta{\tilde\beta} |B|}.
$$

{\it (c) Bound for $\psi^\beta$.} Finally, we estimate the expression involving
$\psi^\beta$. We first observe that 
\be
\label{male}
\e{\alpha \beta} |\psi_A^\beta(g_A)| \leq 1 
\end{equation}
for any $A \subset\bbZ^\nu$ and with $\alpha = \frac12 R^{-2\nu} \Delta_0$,
Indeed,
\bm
\e{\alpha \beta} |\psi_A^\beta(g_A)| = \e{\alpha \beta} |\Psi_A^\beta(g_A) -
\Psi_A(g_A)| = \\
= \e{\alpha \beta} \biggl| -  \int_{\caC_\Lambda^{\sma}} \dd\bsC
\indicator{\bsC \sim g_A,A_C =A,I_C \ni 0, C \subset \Lambda \times
[0,\beta]_\per, |I_C| = \beta} \frac{\Trfunc(\bsC)}{|I_C|}  +\\
+ \int_{\caC_\Lambda^{\sma}} \dd\bsC \indicator{\bsC \sim g_A,A_C =A, I_C \ni
0, C \subset \Lambda \times [-\infty,\infty], |I_C| \geq \beta}
\frac{\Trfunc(\bsC)}{|I_C|}  \biggr| .
\end{multline}
The first integral above corresponds to clusters wrapped around the torus in
vertical direction, while the second one assumes integration over all clusters
in $\Lambda \times [-\infty,\infty]$. For any $\bsC$ above it is $|I_C| \geq
\beta$ and thus
$$
\e{\alpha \beta} \leq \prod_{\xi \in \bsC} \e{\alpha |B|} .
$$
Observing now that every cluster in both integrals necessarily contains in its
support at least one of the points $(x,0)$, $x\in A$, and using the fact that
$\diam A\leq R$, we can bound the first integral by
$$
\frac{R^\nu}\beta \int_{\caC_\Lambda^{\sma}} \dd\bsC \indicator{C\ni (x,0)}
|\Trfunc(\bsC)|  \prod_{\xi \in \bsC} \e{\alpha |B|} ,
$$
which can be directly evaluated by \eqref{klastrodhad}. The same bound can be
actually used also for the second integral, once we realize that the estimate
\eqref{klastrodhad} is  uniform in $\beta$.

Using now  the fact that $\psi^\beta_A = 0$ if $\diam A \geq R$, the condition
$M(A \times \{\tau\}) = M$ implies that $M$ has less than $R^\nu$ sites, hence
$\e{c|M|} \leq \e{c R^\nu}$. Furthermore, referring to \eqref{male}, we have
\be
\int_{\bbT_\Lambda} \dd(A,\tau) |\psi^\beta_A(\boldsymbol\cdot)| \indicator{M(A
\times \{\tau\}) = M} \e{c|M|}\leq \frac{\tilde\beta}\Delta \e{-\frac12
R^{-2\nu} \Delta_0 \beta +c R^\nu} ,
\end{equation}
which can be made small for $\beta$ sufficiently large and concludes thus the
proof of the lemma.

\end{proof}
}

Using Lemma \ref{lemboundrest} and introducing $e_0 = \min_{d \in D} e(d)$,
we can estimate the weight $\frz$ of
the contours in the discrete space of cells.

\begin{lemma}
\label{lemboundweight}
Under the Assumptions \ref{assdiagpot}--\ref{assnoquinst}, for
any
$c <
\infty$, there exist $\beta_0, \tilde\beta_0 < \infty$ and $\varepsilon_0 > 0$ such that
for any  $\beta \geq \beta_0$, $\tilde\beta_0 \leq \tilde\beta < 2\tilde\beta_0$, and
$\normV \leq \varepsilon_0$, one has
$$
|\frz(Y)| \leq \e{-\frac{\tilde\beta}\Delta e_0 |Y|} \e{-c|Y|} 
$$
for any contour $Y$. 
\end{lemma}

{\footnotesize
\begin{proof}
For a given $\Gamma$ (such that $M(\supp\Gamma) \subset \supp Y$) with transitions 
$\{\bsA_1,\dots,\bsA_m\}$ at times $\{\tau_1,\dots,\tau_m\}$,
we define $A(\Gamma)=\cup_{i=1}^m \cup_{x\in A_i}[V(x) \times \tau_i]$,
$\caA = M(A(\Gamma))$, and $\caE \subset \supp Y
\setminus \caA$ to be the set of sites $(x,t)$ such that $\bsn^\Gamma_{V(x)}(\tau)
\notin D_{V(x)}$ for some $(x,\tau) \in C(x, t)$. The latter can be split into two
disjoint subsets,
$\caE =\caE^{\core}\cup\caE^{\soft} $, with  $(x,t) \in\caE^{\core}$ whenever 
$\bsn^\Gamma_{V(x)}(\tau) \notin
G_{V(x)}$ for some $(x,\tau) \in C(x, t)$.
    The
condition
$M(\supp\Gamma) \cup \supp\caM = \supp Y$ in \eqref{deffrz} implies the inequality
$$
\e{c|Y|} \leq \e{c(2R)^\nu |A(\Gamma)|} \e{c|\caE|} \prod_{M \in
\caM} \e{c|M|} .
$$
From definitions  \eqref{deffrz} of $\frz(Y)$ and \eqref{defz} of
$z(\gamma)$, and using Assumption \ref{assPeiprop}, we have
\bm
\label{boundfrz}
\e{c|Y|} |\frz(Y)|  \leq \sum_{\caA \subset \supp Y}
\e{-\frac{\tilde\beta}\Delta e_0 |\supp Y \setminus \caA|} \sum_{\caE \subset
\supp Y \setminus \caA}  \sum_{\caE^{\core} \subset \caE} \e{-(\tilde\beta-c)
|\caE\setminus \caE^{\core}|} \e{-(\frac{\tilde\beta}{\Delta}\frac{\Delta_0}2
(2R)^{-\nu}-c)|\caE^{\core}|} \times\\
\times\int_{\caD_\Lambda^{\per}} \dd\Gamma \indicator{M(A(\Gamma)) = \caA,
M(\core \Gamma)=\caE^{\core}} \prod_{i=1}^m \bigl|
\bra{\bsn^\Gamma_{A_i}(\tau_i-0)} \VAi
\ket{\bsn^\Gamma_{A_i}(\tau_i+0)} \bigr| \e{c(2R)^\nu |A_i|}\times\\
\times  \exp\Bigl\{ -\int_{C(\caA)} \dd(x,\tau) e_x^{\Upsilon}
(\bsn^\Gamma(\tau)) \Bigr\} \e{|\tilde\caR(\Gamma)|} \sum_{\caM, \supp\caM
\subset \supp Y} \prod_{M \in \caM} \bigl| \e{\varphi(M; \Gamma)} - 1 \bigr|
\e{c|M|} .
\end{multline}
All  elements in $\caM$ are different, because it is so in the expansion
\eqref{exprest}. Therefore we have
\ba
\sum_{\caM, \supp\caM \subset \supp Y} \prod_{M \in \caM} \bigl| \e{\varphi(M;
\Gamma)} - 1 \bigr| \e{c|M|} &\leq \sum_{n\geq0} \frac1{n!} \Bigl[ \sum_{M
\subset \supp Y} \bigl| \e{\varphi(M; \Gamma)} - 1 \bigr| \e{c|M|} \Bigr]^n
\nn\\
&\leq \sum_{n\geq0} \frac1{n!} \Bigl[ |Y| \sum_{M \ni (x,t)} \bigl|
\e{\varphi(M; \Gamma)} - 1 \bigr| \e{c |M|} \Bigr]^n
\end{align}
and using Lemma \ref{lemboundrest} this may be bounded by $\e{|Y|}$.

In \eqref{defRA} clusters are small, and they must contain a space-time site
$(x,\tau)$ such that there exists $x'$ with $(x',\tau) \in \core\Gamma$ and
$\dist(x,x') < R$. So we have the bound
$$
|\tilde\caR( \Gamma)| \leq (2R)^\nu |\core\Gamma| \int_{\caC_\Lambda^{\sma}}
\dd\bsC \indicator{{C \ni (x,\tau)}} \bigl| \Trfunc(\bsC) \bigr| ,
$$
since $|\Trfunc(\bsC; \Gamma)| \leq |\Trfunc(\bsC)|$.   Taking now, in Lemma
4.1, the constants $c=\alpha_1=\alpha_2=0$ and 
$\delta=\frac{\Delta_0}{4(2R)^{2\nu}}$, and choosing the corresponding
$\varepsilon_0$, we apply \eqref{klastrodhad} to get, for any $\normV\leq
\varepsilon_0$,   the bound
$$
|\tilde\caR( \Gamma)|  \leq \frac{\Delta_0}4 (2R)^{-\nu} |\core\Gamma| \leq 
\frac{\tilde\beta}{\Delta} \frac{\Delta_0}4 (2R)^{-\nu}|\caE^{\core}| .
$$

Assuming $\tilde\beta \geq c$ and  $\frac{\tilde\beta}{\Delta} \frac{\Delta_0}4
\geq (2R)^{\nu}c$ [\cf  \eqref{vlnkaodhad}], we bound $\e{-(\tilde\beta-c)
|\caE\setminus \caE^{\core}|}\e{-(\frac{\tilde\beta}{\Delta} \frac{\Delta_0}4 
(2R)^{-\nu}-c)|\caE^{\core}|}\leq 1$.

Inserting these estimates into \eqref{boundfrz}, we get
\bm
\e{c|Y|} |\frz(Y)| \leq \e{-\frac{\tilde\beta}\Delta e_0 |Y|} \e{|Y|} \!\!\!
\sum_{\caA \subset \supp Y} \!\!\! \!\!  3^{|\supp Y \setminus \caA|}
\int_{\caD_\Lambda^{\per}}
\dd\Gamma
\indicator{M(A(\Gamma)) =
\caA} \prod_{i=1}^m \bigl|
\bra{\bsn^\Gamma_{A_i}(\tau_i-0)} \VAi
\ket{\bsn^\Gamma_{A_i}(\tau_i+0)} \bigr| \times \\  \times \e{c(2R)^\nu |A_i|}
\exp\Bigl\{ -\int_{C(\caA)} \dd(x,\tau) [e_x^{\Upsilon}
(\bsn^\Gamma(\tau)) - e_0] \Bigr\} .
\end{multline}

To estimate the above expression, we will split the ``transition part'' of the
considered quantum contours  into connected components, to be called {\it 
fragments}, and deal with them separately.  Even though the weight of a quantum
contour cannot be partitioned into the corresponding fragments, we will get an
upper bound  combined from fragment bounds. Consider thus the set
$$
\hat A(\Gamma) = \core \Gamma \cap
C(A(\Gamma)) 
$$
and the fragments  $\zeta_i = (B_i, \bsomega_{B_i})$ on the components
$B_i$ of $\hat A(\Gamma) $,
$\hat A(\Gamma) = \cup_{i=1}^n B_i$, $ \bsomega_{B_i}$ is the restriction of
$\bsomega^\Gamma$ onto $B_i$.

From Assumption \ref{assPeiprop}, we have
$$
\int_{C(\caA)} \dd(x,\tau) \Bigl[ e_x^{\Upsilon}
(\bsn^\Gamma(\tau)) - e_0 \Bigr] \geq \tfrac12 (2R)^{-\nu} \Delta_0
\sum_{i=1}^n |B_i| .
$$
Let us introduce a bound for the contribution of a fragment
$\zeta$ with transitions $\bsA_j, j=1,\dots,k$,
$$
\hat z(\zeta) = \e{-\frac12 (2R)^{-\nu} \Delta_0 |B|} \prod_{j=1}^{k}
|\bra{\bsn^\zeta_{A_j}(\tau_1 - 0)} \VAj \ket{\bsn^\zeta_{A_j}(\tau_1
+ 0)}| \e{c(2R)^\nu  |A_j|} .
$$
Then, integrating over the set $\caF_{C(\caA)}$ of all fragments in $C(\caA)$, we get 
\be
\e{c|Y|} |\frz(Y)| \leq \e{-\frac{\tilde\beta}\Delta e_0 |Y|} \e{|Y|} \sum_{\caA
\subset \supp Y} 3^{|\supp Y \setminus \caA|} \sum_{n \geq 0} \frac1{n!} \Bigl(
\int_{\caF_{C(\caA)}} \dd\zeta \hat z(\zeta) \Bigr)^n .
\end{equation}

Anticipating the bound  $\int_{\caF_{C(\caA)}} \dd\zeta \hat z(\zeta)
\leq |\caA|$, we immediately get the claim, 
$$
\e{c|Y|} |\frz(Y)| \leq \e{-\frac{\tilde\beta}\Delta e_0 |Y|} \e{3|Y|}, 
$$
with a slight change of constant $c \to  c-3$.

{\it A bound on the integral of fragments.} Let us first consider
{\it short} fragments $\zeta = (B, \bsomega_B)$ satisfying the condition 
\be
\label{cond2}
\frac12 \sum_{j=1}^k |A_j| \leq \frac{\log(\Delta / \tilde\beta)}{\log\normV} .
\end{equation}
The integral over the time of  occurrence of the first transition yields the factor
$\tilde\beta/\Delta$. Notice that  $\zeta$ is not a loop.
This follows from the construction of quantum contours and the fact that $B$ is a connected
component of $\hat A(\Gamma)$, where every transition is taken together with its
$R$-neighbourhood.
 Thus,  either its sequence of transitions does not
belong to $\caS$, or the starting configuration does not coincide with the
ending configuration. In the first case we use Assumption \ref{asscompl}, in
the second case Assumption \ref{assnoquinst}, and since \eqref{cond2} means that
the sum over transitions is bounded, we can write
\be
\int_{\caF_{C(\caA)}^{\operatorname{short}}} \dd\zeta \hat z(\zeta)  \leq
\tfrac12 |\caA| .
\end{equation}

Finally, we estimate the integral over $\zeta$'s that are not short. 
We have
\be
\int_{\caF_{C(\caA)}\setminus\caF_{C(\caA)}^{\operatorname{short}}} \dd\zeta
\hat z(\zeta) \leq  |\caA| \frac{\tilde\beta}\Delta
\int_{\caF_{C(\caA)}^{(x,\tau)}\setminus\caF_{C(\caA)}^{\operatorname{short}}}\dd\zeta
\hat z(\zeta).
\end{equation}
Here $\caF_{C(\caA)}^{(x,\tau)}$ is the set of all fragments $\zeta$ whose 
first quantum transition $(A_1, \tau_1)$ is such that $x\in A_1$ and 
$\tau=\tau_1$.  Whenever $\zeta$ is not short, we have
$$
1 \leq \frac\Delta{\tilde\beta} \prod_{j=1}^k \normV^{-\frac12 |A_j|} .
$$
Thus, defining
\be
\hat z'(\zeta) = \e{-\frac12 (2R)^{-\nu} \Delta_0 |B|} \prod_{j=1}^k 
 \Bigl[\normV^{\frac12} \e{c(2R)^\nu +1} \Bigl]^{|A_j|} ,
\end{equation}
we find the bound
$$
|\caA| \int_{\caF(x,\tau)}\dd\zeta \hat z'(\zeta) .
$$
Here, slightly overestimating, we take for $\caF(x,\tau)$ the set of all
fragments containing a quantum transition $(A,\tau)$ with $x\in A$.

The support $B$ of a  fragment $\zeta = (B,\bsomega_B)\in\caF(x,\tau)$, 
 is a finite union of
vertical segments (\ie sets of the form $\{y\} \times [\tau_1,
\tau_2] \subset
\bbT_\Lambda$) and $k$ horizontal quantum transitions $A_1,\dots,A_k$. 

We will finish the proof by proving by induction the bound
\be
\label{indukce}
\int_{\caF(x,\tau; k)}\dd\zeta \hat z'(\zeta) \le 1
\end{equation}
with $\caF(x,\tau; k)$ denoting the set of fragments from $\caF(x,\tau)$
with at most $k$ quantum transitions.

Consider thus a fragment $\zeta$ with $k$ horizontal quantum transitions
connected by vertical segments. Let $(A,\tau)$ be the transition containing 
the point $(x,\tau)$ and let $(A_1,\tau+\tau_1), \dots ,
(A_\ell,\tau+\tau_\ell)$ be the transitions that are connected by (one or
several) vertical segments of the respective lengths $|\tau_1|, \dots,
|\tau_\ell|$ with the transition $(A,\tau)$. If we remove all those segments,
the fragment $\zeta$ will split into the ``naked'' transition $(A,\tau)$ and
additional  $\bar\ell\le \ell$ fragments $\zeta_1,\dots , \zeta_{\bar\ell}$,
such that each  fragment $\zeta_j$, $j=1, \dots , \bar\ell$, belongs to  
$\caF(y_j,\tau+\tau_j; k-1)$ with $y_j\in A$. Taking into account that the
number of configurations (determining the possible vertical segments attached
to $A$) above and below $A$ is bounded by $S^{2|A|}$ and that the number of
possibilities to choose the points $y_j$ is bounded by $|A|^{\bar\ell}$, we get
\bm
\int_{\caF(x,\tau; k)} \dd\zeta \hat z'(\zeta)  \leq \\
\leq \sum_{A, \dist(A,x) < R} \bigl[ \normV^{\frac12} \e{c(2R)^\nu +1}S^2
\bigr]^{|A|} \sum_{\bar\ell=1}^{\infty} \frac{|A|^{\bar\ell}}{\bar\ell !} \int
\dd\tau_1\dots\!\int \dd\tau_{\bar\ell}\,\e{-\frac12 (2R)^{-\nu}\Delta_0
(\tau_1+\dots +\tau_{\bar\ell})}  \prod_{j=1}^{\bar\ell}
\int_{\caF(y_j,\tau+\tau_j; k-1)} \dd\zeta \hat z'(\zeta_j) \\
\leq \sum_{A, \dist(A,x) < R} \bigl[ \normV^{\frac12} S^2 \e{c(2R)^\nu +2}\bigr]^{|A|}
\e{2(2R)^\nu / \Delta_0}  \leq 1
\end{multline}
once $\normV$ is sufficiently small.
\end{proof}
}

In the application of Pirogov-Sinai theory we shall also need a bound on
derivatives of the weight of contours. 

\begin{lemma}
Under the Assumptions \ref{assdiagpot}--\ref{assboundsder}, for any $c <
\infty$, there exist constants  $\alpha, \beta_0, \tilde\beta_0 < \infty$ and $\varepsilon_0 >
0$ such that if $\beta \geq \beta_0$, $\tilde\beta_0 \leq \tilde\beta <
2\tilde\beta_0$, and $\normV + \sum_{i=1}^{r-1} \| \frac\partial{\partial\mu_i}
\bsV\| \leq \varepsilon_0$, one has
$$
\bigl|\frac\partial{\partial\mu_i} \frz(Y)\bigr| \leq \alpha \tilde\beta |Y|
\e{-\frac{\tilde\beta}\Delta e^\mu_0 |Y|} \e{-c|Y|}
$$
for any contour $Y$.
\end{lemma}

{\footnotesize
\begin{proof}
From the definition \eqref{deffrz} of $\frz$, one has
\bm
\bigl|\frac\partial{\partial\mu_i} \frz(Y)\bigr| \leq |\frz(Y)|  \biggl\{ \sum_{\gamma
\in \Gamma} \bigl|\frac\partial{\partial\mu_i} z(\Gamma)\bigr| + \sum_{d \in D} \bigl|W_d
\cap C(\supp Y)\bigr| \bigl|\frac\partial{\partial\mu_i} e^\mu(d)\bigr| +
\bigl|\frac\partial{\partial\mu_i} \tilde\caR(\Gamma)\bigr| \biggr\} \\
+ \int_{\caD_\Lambda^{\per}(Y)}\dd\Gamma  \prod_{\gamma
\in \Gamma} |z(\gamma)| \prod_{d \in D} \e{-e^\mu(d) |W_d \cap C(\supp Y)|}
\e{|\tilde\caR(\Gamma)|} \\
\sum_\caM \indicator{\caM(\supp \Gamma) \cup \supp\caM = \supp
Y} \sum_{M \in \caM} \Bigl| \e{\varphi(M; \Gamma)} \frac\partial{\partial\mu_i}
\varphi(M; \Gamma) \Bigr| \prod_{M' \in \caM, M' \neq M} \bigl|
\e{\varphi(M';\Gamma)} -1 \bigr| .
\end{multline}
The bound for $|\frac\partial{\partial\mu_i} z(\Gamma)|$ is standard, see
\cite{BKU}, and $|\frac\partial{\partial\mu_i} e^\mu(d)|$ is assumed to be
bounded in Assumption \ref{assboundsder}. For the other terms we have to control
clusters of loops. Since we have exponential decay for $z(\xi)$ with any
strength (by taking $\beta$ large and $\normV$ small), we have the same for
$\frac\partial{\partial\mu_i} z(\xi)$ (by taking $\beta$ larger and $\normV$
smaller). The integrals over $\bsC$ can be estimated as before, the only effect
of the derivative being an extra factor $n$ (when the clusters have $n$ loops).
\end{proof}
}

\section{Expectation values of local observables and construction of pure states}
\label{secexpobs}

So far we have obtained an expression \eqref{periodparticni}
for the partition function
$Z_\Lambda^{\per}$ of the quantum model on torus $\Lambda$ in terms of that of a classical
lattice contour model  with the weights of the contours showing an exponential decay with
respect to their length.  Using the same weights $\frz(Y)$, we can also introduce the
partition functions
$Z_{\Lambda(L)}^d$ with the torus $\Lambda$ replaced by a hypercube $\Lambda(L)$
and with  fixed boundary conditions
$d$. Namely, we take simply the sum  only over those collections $\caY $ of contours
whose external contours are labeled by $d$ and are not close to the boundary%
\footnote{In the terminology of Pirogov-Sinai theory we  rather  mean {\it diluted partition
functions} --- see the more precise definition below.} . 
Notice,  however,  that here we are defining $Z_{\Lambda(L)}^d$  directly in terms of the
classical contour model, without ensuring existence of corresponding partition function
directly for the original model.  We will use these partition functions only as a tool for
proving our Theorems that are stated directly in terms of quantum models.

To be more precise,
we can extend the definition even more and consider,  instead of the torus $\Lambda$, 
any finite set
$V\subset\bbL =\bbZ^{\nu}\times\{0,1,\dots, N-1\}^{\per}$.
 There is a class of  contours 
that can be viewed as having their support contained  in $V\subset\bbL $.
 For any such contour $Y$ we
introduce its  interior $\Int Y$ as the union of all finite
components of $\bbL \setminus\supp Y$
and $\Int_d Y$  as the union of all
components of $\Int Y$ whose boundary is labeled by $d$.
Recalling that we assumed $\nu\geq 2$, we
note that the set
$\bbL\setminus (\supp  Y \cup \Int  Y)$
is a connected set, implying that the
label $\alpha_Y(\cdot)$ is constant on
the boundary of the set $V(Y)=\supp Y\cup\Int Y$.
We say that $Y$ is a $d$-contour, if $\alpha_Y = d$ on this boundary.
Two contours $Y$ and $Y^\prime$ are
called  {\it mutually external}
if $V(Y)\cap V(Y^\prime)=\emptyset$.
Given an admissible set $\caY$ of contours,
 we say that $Y\in\caY$ is an {\it external contour} in $\caY$,
if $\supp Y\cap V(Y^\prime)=\emptyset$
for all $Y^\prime\in\caY$, $Y^\prime\neq Y$.
The sets $\caY$ contributing to  $Z_{V}^d$
are such that all their external contours are $d$-contours and 
$\dist(Y, \partial V)> 1$ for every $Y\in \caY$.

In this way we found ourselves exactly in the setting of standard Pirogov-Sinai theory, or
rather, the reformulation for ``thin slab'' (cylinder $\bbL$ of fixed temporal
size $N$) as presented in Sections 5--7 and  Appendix of \cite{BKU}.
In particular,  for sufficiently large $\beta$ and sufficiently small 
$\normV +
\sum_{i=1}^{r-1} \| \frac\partial{\partial\mu_i} \bsV\| $,  there exist  functions
$f^{\beta,\mu}(d)$, metastable free energies,  such that the condition 
 $\Re f^{\beta,\mu}(d) = f_0$, with $f_0\equiv f_0^{\beta,\mu}$ defined by $f_0= \min_{d' \in D} \Re
f^{\beta,\mu}(d')$, characterizes   the existence of pure stable phase $d$.
Namely, as will be shown  next, a  pure stable phase 
$\expval{\boldsymbol\cdot}_\beta^{d}$  exists and is  close to the  pure ground state
$\ket{d}$.

There is one subtlety in the definition of $f^{\beta,\mu}(d)$. Namely, after choosing a
suitable
$\tilde\beta_0$, given $\beta$, there exist several pairs $(\tilde\beta,N)$ such that
$\tilde\beta\in(\tilde\beta_0, 2\tilde\beta_0)$ and $N\tilde\beta=\beta$.
To be specific,
we may agree to choose among them that one with maximal $N$. The function
$f^{\beta,\mu}(d)$ is then uniquely defined for each $\beta>\beta_0$. Notice, however, that
while increasing $\beta$, we pass, at the particular value $\beta_N=N\tilde\beta_0$, from
discretization of temporal size $N$ to $N+1$.
As a result, the function $f^{\beta,\mu}(d)$ might be discontinuous at $\beta_N$ with
$\beta=\infty$ being an accumulation point of such discontinuities. Nevertheless, these
discontinuities are harmless. They can appear only when $\Re f^{\beta,\mu}(d)>f_0$ and do not
change anything in the following argument.

Before we come to the construction of pure stable phases, notice that the first claim
of Theorem \ref{thmphases} (equality of $f_0$ with the limiting free energy) is now a direct
consequence of the bound  
\be
\Bigl|
  Z_\Lambda^{\per} - |Q|  e^{-\tilde\beta f_0 N L^\nu}\Bigr|
\leq e^{-\tilde\beta f_0 N L^\nu}O(e^{-\text{const } c\, L})
\end{equation}
(\cf \cite{BKU}, (7.14)). Here  $Q=\{d; \Re f^{\beta,\mu}(d) =f_0 \}$.

The expectation value of a local observable $T$ is defined as
\be
\expval T_\Lambda^{\per} = \frac{\Tr T \e{-\beta
\bsH_\Lambda^{\per}}}{\Tr
\e{-\beta \bsH_\Lambda^{\per}}}  .
\end{equation}
In Section \ref{secDuhamexp} we have obtained a contour expression for
$Z_\Lambda^{\per} = \Tr \e{-\beta \bsH_\Lambda^{\per}}$. We retrace here  the same steps
for $Z_\Lambda^{\per}(T) \isdefby \Tr T \e{-\beta \bsH_\Lambda^{\per}}$. 
Duhamel
expansion \eqref{Duhamexp} for $Z_\Lambda^{\per}(T)$ leads to an equation analogous
to \eqref{Zexp},
\bm
Z_\Lambda^{\per}(T) = \sum_{m \geq 0} \sum_{n_\Lambda^0, \dots n_\Lambda^m}
\sumtwo{\bsA_1, \dots, \bsA_m}{\bar A_i \subset \Lambda}
\int_{0<\tau_1<...<\tau_m<\beta} \dd\tau_1 \dots \dd\tau_m \bra{n_\Lambda^0} T
\ket{n_\Lambda^1} \\ 
\e{-\tau_1 \tilde H^{(0)\per}_\Lambda(n_\Lambda^1)} \bra{n_\Lambda^1}
\VAone \ket{n_\Lambda^2} \e{-(\tau_2-\tau_1) \tilde
H^{(0)\per}_\Lambda(n_\Lambda^2)} \dots \bra{n_\Lambda^m} \VAm
\ket{n_\Lambda^0} \e{-(\beta-\tau_m) \tilde H^{(0)\per}_\Lambda(n_\Lambda^0)} .
\end{multline}
Configurations $n^0_\Lambda$ and $n^1_\Lambda$ match on $\Lambda \setminus
\supp T$ ($\supp T \subset \Lambda$ is a finite set due to the locality of $T$),
but may differ on $\supp T$ if $T$ is an operator with non zero off-diagonal
terms. Let $\caQ_{\Lambda}^{\per}(T)$ be the set of quantum configurations with
$\bsn_{\Lambda}(\tau)$ that is constant
except possibly at $\cup_{i=1}^m (A_i \times \tau_i) \cup (\supp T \times 0)$.
Then
\be
Z_\Lambda^{\per}(T) = \int_{\caQ_{\Lambda}^{\per}(T)} \dd\bsomega_{\bbT_\Lambda}
\bra{n_\Lambda^0} T \ket{n_\Lambda^1} \rho^{\per}(\bsomega_{\bbT_\Lambda}) .
\end{equation}

We identify loops with the same iteration scheme as in Section
\ref{secDuhamexp}, starting with the set $ \bsB^{(0)}(\bsomega) \union (\supp T
\times 0) $ instead of $\bsB^{(0)}(\bsomega)$ only. This leads to the set
$\bsB^T(\bsomega)$. Removing the loops, we define $\bsB^T_{\rm e}(\bsomega)$,
whose connected components  form quantum contours. There is one special quantum
contour, namely that which contains $\supp T \times 0$. Let us denote it by
$\gamma^T$ and define its weight [see \eqref{defz}]
\bm
z^T(\gamma^T) = \bra{\bsn^{\gamma^T}_{\supp T}(-0)} T
\ket{\bsn^{\gamma^T}_{\supp T}(+0)} \prod_{i=1}^m
\bra{\bsn_{A_i}^{\gamma^T}(\tau_i-0)} \VAi
\ket{\bsn_{A_i}^{\gamma^T}(\tau_i+0)} \\
\exp \Bigl\{ -\int_B \dd(x,\tau) 
e_x^{\Upsilon}(\bsn^{\gamma^T}(\tau)) 
\Bigr\}.
\end{multline}
Let $\Gamma^T = \{ \gamma^T, \gamma_1, \dots, \gamma_k \}$ be an admissible set
of quantum contours, defining a quantum configuration $\bsomega^{\Gamma^T} \in
\caQ_{\Lambda}^{\per}(T)$. Then we have an expression similar to that of Lemma
\ref{lemfpart},
\be
Z_\Lambda^{\per}(T) = \int_{\caD_\Lambda^{\per}(T)}\dd\Gamma^T \prod_{d \in D}
\e{-|W_d(\Gamma^T)| e(d)} z^T(\gamma^T) \prod_{\gamma \in \Gamma^T \setminus \{
\gamma^T \} } z(\gamma)
\e{\caR(\Gamma^T)} ,
\end{equation}
with $\caR(\Gamma^T)$ as in \eqref{deffrR} with $\Gamma$ replaced
by $\Gamma^T$.

Next step is to discretize the lattice, to expand $\e{\caR(\Gamma^T)}$, and if
$Y^T$ is the contour that contains $\supp T \times 0 \subset \bbL_\Lambda$, to
define $\frz^T(Y^T)$ [see \eqref{deffrz}]:
\bm
\frz^T(Y^T) = \int_{\caD_\Lambda^{\per}(Y^T)}\dd\Gamma^T
 z^T(\gamma^T) \prod_{\gamma \in \Gamma^T \setminus \{
\gamma^T \} } z(\gamma) \prod_{d \in D} \e{-e(d) |W_d(\Gamma^T) \cap C(\supp
Y^T)|} \e{\tilde\caR(\Gamma^T)} \\
\sum_\caM \indicator{M(\supp\Gamma^T \union \supp\caM =
\supp Y^T} \prod_{M \in \caM} \bigl( \e{\varphi(M; \Gamma^T)} - 1 \bigr) .
\end{multline}

We also need a bound for $\frz^T(Y^T)$. It is clear that the situation is the
same as for Lemmas \ref{lemboundrest} and \ref{lemboundweight}, except for a
factor $\bra{\bsn^{\gamma^T}_{\supp T}(-0)} T \ket{\bsn^{\gamma^T}_{\supp
T}(+0)}$ that is bounded by $\| T \|$. We can thus summarize:

\begin{lemma}
Under the Assumptions \ref{assdiagpot}--\ref{assnoquinst}, for any $c <
\infty$, there exist $\beta_0, \tilde\beta_0 < \infty$, and $\varepsilon_0 > 0$
such that if $\beta \geq \beta_0$, $\tilde\beta_0 \leq \tilde\beta <
2\tilde\beta_0$ and $\normV \leq \varepsilon_0$, we have
\be
\label{pozorovatelna}
Z_\Lambda^{\per}(T)  = \sum_{\caY^T = \{ Y^T, Y_1, \dots, Y_k \} } \prod_{d \in D}
\e{-\frac{\tilde\beta}\Delta e(d) |\caW_d(\caY^T)|} \frz^T(Y^T) \prod_{Y \in
\caY^T \setminus \{ Y^T \} } \frz(Y) ,
\end{equation}
for every local observable $T$, with 
$$
|\frz^T(Y^T)| \leq \| T \| \e{c|\supp T|} \e{-\frac{\tilde\beta}\Delta e_0
|Y^T|} \e{-c|Y^T|} 
$$
for any contour $Y^T$.
\end{lemma}

In a similar manner as at the beginning of this section, we can introduce 
$Z_{V}^d(T)$ for any $V\subset\bbL$ by restricting ourselves in the sum
\eqref{pozorovatelna} to the collections $\caY^T$ whose all  external contours are
$d$-contours and 
$\dist(Y, \partial V)> 1$ for every $Y\in \caY^T$.
Thus we can {\it define} the expectation value
\be
\expval T_{V}^d = \frac{Z_{V}^d(T)}{Z_{V}^d}  
\end{equation}
for any $V\subset\bbL$ and, in particular,  the expectation $\expval T_{\Lambda(L)}^d$
for a hypercube $\Lambda(L)$.

Again, this is exactly the setting discussed in detail in \cite{BKU}.
We can use directly the corresponding results (\cf \cite{BKU}, Lemma 6.1) to prove first
that the limiting state  $\expval{\boldsymbol\cdot}_\beta^{d}$ exists. 
Further, retracing the proof of Theorem 2.2 in \cite{BKU} we prove that the limit
\be
\expval T_\beta^{\per} =\lim_{\Lambda\nearrow \bbZ^\nu} \frac{\Tr T \e{-\beta
\bsH_\Lambda^{\per}}}{\Tr
\e{-\beta \bsH_\Lambda^{\per}}} 
\end{equation}
exists for every local $T$ (proving thus Theorem \ref{thmlimit}).
Moreover,
\be
\expval T_\beta^{\per} =\frac1Q\sum_{d\in Q}\expval{T}_\beta^{d}, 
\end{equation}
where, again, $Q$ denotes the set of stable phases, $Q=\{d; \Re f^{\beta,\mu}(d) =f_0 \}$.
Thus we proved the claim d) of Theorem \ref{thmphases}.

Also the assertion c) follows in standard manner from contour representation employing directly  the
exponential decay of contour activities and corresponding cluster expansion (\cf \cite{BKU}, (2.27)). 

Before passing to the proof of b),  we shall verify that $\expval{\boldsymbol\cdot}_\beta^d$
is actually a pure stable state according to our definition, \ie a limit of unperturbable
states%
\footnote{Recall that, up to now, the state $\expval{\boldsymbol\cdot}_\beta^d$ is defined only in terms
of the contour representation (see (6.9), (6.8), and (4.34)), and the only proven connection with a state of
original quantum model is the equality (6.11).}.  To this end, let us first discuss how metastable free energies 
$f^{\beta,\mu}(d)$ change with $\mu$. The standard construction yields $f^{\beta,\mu}(d)$
in the form of a sum $e^{\mu}(d)+s^{\beta,\mu}(d)$, where $s^{\beta,\mu}(d)$ is the free
energy of ``truncated'' contour model $K_d^{\prime}(Y)$ (see \cite{BKU}, (5.13) and (5.6))
constructed from labeled contour model (4.34), which is  under control by cluster
expansions. As a result, we have bounds of the form $O\bigl(e^{-\beta}+\normV +
\sum_{i=1}^{r-1} \Bigl\| \frac{\partial\bsV}{\partial\mu_i} \Bigr\| \bigr)$ on
$|s^{\beta,\mu}(d)|$  as well as on the  derivatives with respect to  $\mu$.
Hence, in view of Assumption 7, the leading behaviour is yielded by $e^\mu(d)$.

Starting thus from a given potential $\Phi^\mu$ with 
$Q^\mu=\{d\in D; \Re f^{\beta,\mu}(d) =f_0^\mu\}$, one can easily add to $\Phi^\mu$ a
suitable ``external field'' that favours a chosen $d\in Q^\mu$.  For example, one can take
$$
\Phi^{\mu,\alpha}_A(n) = \Phi^{\mu}_A(n)+ \alpha \delta_A^d(n)
$$
with $\delta_A^d$ defined by taking $\delta_A^d(n)=0$ for $n_A=d_A$ and  
$\delta_A^d(n)=1$ otherwise%
\footnote{Actually, we can restrict $\delta_A^d$  only to a particular type of sets $A$ ---
for example all hypercubes of side $R$.}.
Now, since $\frac{\partial e^{\mu,\alpha}(d)}{\partial\alpha}$ is bounded from below  by a positive constant 
(while $\frac{\partial e^{\mu,\alpha}(d')}{\partial\alpha}=0$ for $d'\neq d$), for any
$\alpha>0$ the only stable phase is $d$,
$\Re f^{\beta,\mu, \alpha}(d) =f_0^{\beta,\mu,\alpha}\equiv \min_{d'\in D}\Re f^{\beta,\mu,
\alpha}(d') $, and, in the same time, $\Re f^{\beta,\mu, \alpha}(d') >f_0^{\beta,\mu,\alpha}$
for  $d'\neq d$. Thus, $Q^{\mu,\alpha}=\{d\}$ and 
$\expval{\boldsymbol\cdot}_{\beta,\mu,\alpha}^d=
\expval{\boldsymbol\cdot}_{\beta,\mu,\alpha}^{\per}$. This state is unperturbable ---
when adding any small perturbation, metastable free energies will change only a little and
that one corresponding to the state $d$ will still be the only one attaining the minimum.
The fact that in the limit of vanishing perturbation we recover
$\expval{\boldsymbol\cdot}_{\beta,\mu,\alpha}^d$, as well as the fact that 
$$
\lim_{\alpha\to 0+}\expval{\boldsymbol\cdot}_{\beta,\mu,\alpha}^{\per}
\equiv \lim_{\alpha\to 0+}\expval{\boldsymbol\cdot}_{\beta,\mu,\alpha}^d
=\expval{\boldsymbol\cdot}_{\beta,\mu}^d,
$$
follows by inspecting the contour representations of the corresponding expectations and
observing that it can be expressed in terms of converging cluster expansions whose terms
depend smoothly on $\alpha$ as well as on the additional perturbation.

To prove, finally, the claim b) of Theorem \ref{thmphases}, it suffices to show that it is
valid for 
$\expval{\boldsymbol\cdot}_{\beta,\mu,\alpha}^{\per}
=\expval{\boldsymbol\cdot}_{\beta,\mu,\alpha}^d$ for every $\alpha>0$.
Abbreviating $\expval{\boldsymbol\cdot}_{\beta,\mu,\alpha}^{\per}=
\expval{\boldsymbol\cdot}^{\per}$ and $\bsH_\Lambda^{\mu,\alpha,\per}=
\bsH_\Lambda^{\per}$, we first notice that
the expectation value of the projector onto  the configuration  $d$ on $\supp T$,
$P^d_{\supp T} \isdefby \ket{d_{\supp T}} \bra{d_{\supp T}}$, is close to 1, since its
complement $\expval{(\bbbone - P^d_{\supp T})}^{\per}=\expval{(\bbbone - P^d_{\supp
T})}^d$ is related to the presence of a contour intersecting or surrounding $\supp T$ (loops
intersecting $\supp T \times \{0\}$ are considered here as part of quantum contours),
whose weight is small. More precisely, for any $\delta > 0$ we have
$$
\expval{(\bbbone - P_{\supp T}^d)}^{\per} \leq \delta |\supp T| ,
$$
whenever  $\normV$ is small enough
and $\beta$ large enough.
Furthermore,
\bm
\expval T_\Lambda^{\per}= \frac1{Z_\Lambda^{\per}} \Bigl[ \Tr \bigl(P^d_{\supp T} T P^d_{\supp
T} \e{-\beta \bsH_\Lambda^{\per}}\bigr)+\\ +\Tr\bigl( (\bbbone - P^d_{\supp T}) T P^d_{\supp
T} \e{-\beta \bsH_\Lambda^{\per}}\bigr) + \Tr\bigl( T (\bbbone - P^d_{\supp T}) \e{-\beta
\bsH_\Lambda^{\per}} \bigr)\Bigr]
\end{multline}
and
\ba
\Tr \bigl(P^d_{\supp T} T P^d_{\supp T} \e{-\beta \bsH_\Lambda^{\per}}\bigr) &=
\bra{d_\Lambda} T \ket{d_\Lambda} \Tr\bigl( P^d_{\supp T} \e{-\beta
\bsH_\Lambda^{\per}} \bigr)\nn\\
&= \bra{d_\Lambda} T \ket{d_\Lambda} \bigl[ \Tr \bigl(\e{-\beta\bsH_\Lambda^{\per}}\bigr)
- \Tr \bigl( (\bbbone - P^d_{\supp T}) \e{-\beta \bsH_\Lambda^{\per}} \bigr)\bigr] ,
\end{align}
so that we have
\bm
\bigl| \expval T_\Lambda^{\per} - \bra{d_\Lambda} T \ket{d_\Lambda} \bigr| \leq\\
\leq
\bigl| \bra{d_\Lambda} T \ket{d_\Lambda} \bigr| \expval{(\bbbone - P^d_{\supp
T})}_\Lambda^{\per} + \bigl| \expval{(\bbbone - P^d_{\supp T}) T P^d_{\supp
T}}_\Lambda^{\per} \bigr| + \bigl| \expval{T (\bbbone - P^d_{\supp T})}_\Lambda^{\per}
\bigr|.
\end{multline}
The mapping $(T,T') \mapsto \expval{T^\dagger T'}_\Lambda^{\per}$, with  any two
local operators
$T,T'$, is a scalar
product; therefore the Schwarz inequality  yields
\bm
\bigl| \expval T_\Lambda^{\per} - \bra{d_\Lambda} T \ket{d_\Lambda} \bigr| \leq
\bigl| \bra{d_\Lambda} T \ket{d_\Lambda} \bigr| \expval{(\bbbone - P^d_{\supp
T})}_\Lambda^{\per}\\
\hfill + \Bigl( \expval{(\bbbone - P^d_{\supp T})}_\Lambda^{\per}\Bigr)^{\frac12}
\Bigl( \bigl[ \expval{P^d_{\supp T} T^\dagger T P^d_{\supp T}}_\Lambda^{\per}
\bigr]^{\frac12} + \bigl[ \expval{T^\dagger T}_\Lambda^{\per} \bigr]^{\frac12}
\Bigr) \\
\leq \| T \| \Bigl[ \expval{(\bbbone - P_{\supp T}^d)}_\Lambda^{\per} + 2 \bigl(
\expval{(\bbbone - P_{\supp T}^d)}_\Lambda^{\per} \bigr)^{1/2} \Bigr] 
\leq \| T \| |\supp T| (\delta + 2 \delta^{\frac12}).
\end{multline}

The proof of the remaining Theorem 2.3 is  a standard application of the implicit function theorem.
Thus, for example, the point $\bar\mu_0$ of maximal coexistence,
$\Re f^{\beta,\bar\mu_0}(d)=\Re f^{\beta,\bar\mu_0}(d')$ for every pair $d,d'\in D$,
can be viewed as the solution of the vector equation $f(\bar\mu_0)=0$, with 
$f(\mu)=(\Re f^{\beta,\mu}(d_i)-\Re f^{\beta,\mu}(d_r))_{i=1}^{r-1}$.
Now, $f=e+s$, $e(\mu)=(e^{\mu}(d_i)-e^{\mu}(d_r))_{i=1}^{r-1}$,
$s(\mu)=(\Re s^{\beta,\mu}(d_i)-\Re s^{\beta,\mu}(d_r))_{i=1}^{r-1}$, with
$\|s\|$ as well as $\bigl\|\frac{\partial s}{\partial \mu}\bigr\|$
bounded by a small constant once $\normV +
\sum_{i=1}^{r-1} \Bigl\| \frac{\partial\bsV}{\partial\mu_i} \Bigr\|$ is sufficiently small
$\beta$ is sufficiently large. 
The existence of a unique solution $\bar\mu_0\in\caU$ then follows once we notice the
existence of the solution $\mu_0\in\caU$ of the equation $e(\mu_0)=0$ (equivalent with 
$e^{\mu_0}(d)=e^{\mu_0}(d')$, $d,d'\in D$) and the fact that the mapping
$$
\caT: \mu\to A^{-1}\bigl( {\frac{\partial
e}{\partial\mu}}\bigr|_{\mu=\mu_0}(\mu-\mu_0)-f(\mu)\bigr)
$$
with $A^{-1}$ the matrix inverse to $\bigl({\frac{\partial
e}{\partial\mu}}\bigr)$, is a contraction. To this end it is enough just to recall Assumption 7
and the bounds on $s^{\beta,\mu}(d)$, $d\in D$,  and its derivatives.

\end{document}